\documentclass[12pt]{article}
\usepackage{array}
\usepackage{epsfig}

\makeatletter
\def\alt{\mathrel{\mathpalette\gl@align<}}
\def\agt{\mathrel{\mathpalette\gl@align>}}
\def\gl@align#1#2{\lower.6ex\vbox{\baselineskip\z@skip\lineskip\z@
\ialign{$\m@th#1\hfil##\hfil$\crcr#2\crcr\sim\crcr}}}
\makeatother

\begin{document}
\begin{flushright}
{\tt hep-ph/0504082}\\
OSU-HEP-05-05\\
April, 2005 \\
\end{flushright}
\vspace*{1.0cm}
\begin{center}
{\baselineskip 25pt
\Large{\bf 
Coupling Unifications in Gauge-Higgs Unified \\[2.5mm] Orbifold Models
}}

\vspace{1cm}

{\large
Ilia Gogoladze$^a$\footnote
{email: {\tt gogoladze.1@nd.edu}}, 
Tianjun Li$^b$\footnote
{email: {\tt tli@ias.edu}},
Yukihiro Mimura$^c$\footnote
{email: {\tt mimura2y@uregina.ca}}
and
S. Nandi$^d$\footnote
{email: {\tt shaown@okstate.edu}}
}
\vspace{.5cm}

{\small {\it 
$^a$Department of Physics, University of Notre Dame, Notre Dame, IN 46556, USA \\
$^b$School of Natural Sciences, Institute for Advanced Study, \\
             Einstein Drive, Princeton, NJ 08540, USA \\
$^c$Department of Physics, University of Regina, Regina, Saskatchewan S4S 0A2, Canada\\
$^d$Physics Department, Oklahoma State University, Stillwater, OK 74078 USA}}


\vspace{1.0cm}
{\bf Abstract}

\end{center}

\baselineskip 16pt

Supersymmetric gauge theories, in higher dimensions compactified in an orbifold, give a natural
framework to unify the gauge bosons, Higgs fields and even the matter fields in a
single multiplet of the unifying gauge symmetry.
The extra dimensions and the supersymmetry
are the two key ingredients for such an unification. In this work, we investigate
various scenarios for the unification of the three gauge couplings, and the Yukawa couplings
in the Minimal Supersymmetric Standard Model (MSSM), as well as the trilinear Higgs couplings
$\lambda$ and $\kappa$ of the Non-Minimal Supersymmetric Standard Model (NMSSM). We present 
an $SU(8)$ model in six dimensions with $N=2$ supersymmetry, compactified in a $T^2/Z_6$ 
orbifold which unifies the three gauge couplings with $\lambda$ and $\kappa$ of NMSSM. Then,
we present an $SU(9)$ model in 6D, which, in addition, includes partial unification of 
Yukawa couplings, either for the up-type (top quark and Dirac tau-neutrino)
or down-type (bottom quark and tau lepton).
We also study 
the phenomenological implications of these various unification scenarios using the
appropriate renormalization group equations, and show that such unification works very
well with the measured low energy values of the couplings. 
The predicted upper bounds for the lightest neutral Higgs boson mass in our model
is higher than those in MSSM, but lower that those in the general NMSSM (where the couplings
$\lambda$ and $\kappa$ are arbitrary). Some of the predictions of our models can be tested in the 
upcoming Large Hadron Collider.

\thispagestyle{empty}

\bigskip
\newpage

\addtocounter{page}{-1}

\section{Introduction}
\baselineskip 18pt

The Standard Model (SM) is well established to describe the physics below the weak scale.
The key ingredient of the electroweak theory is the Higgs mechanism,
in which the electroweak gauge symmetry, $SU(2)_L \times U(1)_Y$,
is spontaneously broken down to the electromagnetic gauge symmetry, $U(1)_{EM}$,
by the vacuum expectation values (VEVs) of Higgs doublets.
The VEVs of Higgs doublets not only do make the $W$ and $Z$ bosons massive,
but also give masses to the quarks and leptons through Yukawa couplings.
Although the Higgs bosons have not been observed yet, it is expected that
there is rich physics in the Higgs sector which will be tested at the
upcoming colliders.
The elementary particles, which we have already observed, are the SM 
fermions and gauge bosons,
while the Higgs particles are scalar fields. 
The radiative corrections to the Higgs boson (or scalars in general)
masses are quadratically
dependent on the UV cutoff scale $\Lambda$, and are not protected
by chiral or gauge symmetries. Thus, the natural Higgs masses are of
order $\Lambda$ rather than the weak scale, leading to the gauge hierarchy
problem because $\Lambda$ should be around the Planck or string
scale. It is expected that there exists new physics around
 a TeV scale from naturalness of the Higgs masses. A well-known solution to
the gauge hierarchy problem is supersymmetry.
In supersymmetric theories, each particle has a superpartner which differs
in spin by $1/2$ and is related to the original particle by a
supersymmetry transformation. Since supersymmetry relates the
scalar and fermionic sectors, the chiral symmetries, which protect
the fermion masses, also protect the scalar masses against the
quadratic divergences, leading to an elegant solution
to the gauge hierarchy problem.
As a result, there are only logarithmic divergences in the 
scalar masses and couplings in supersymmetric theories
and they can be extrapolated to the higher energy using the evolution
of the Renormalization Group Equations (RGEs).
In fact, in the Minimal Supersymmetric
Standard Model (MSSM), the three gauge couplings in the SM are unified
at a scale of about $2\times 10^{16}$ GeV~\cite{Amaldi:1991cn},
 which implies a possible Grand Unified Theory (GUT).
In addition to the gauge coupling unification,
 the Yukawa couplings can be unified
(bottom-tau~\cite{Komine:2001rm}, top-bottom-tau~\cite{Baer:2001yy}, 
or other types of Yukawa unification~\cite{Fukuyama:2002ch}).
In such a way, the supersymmetric theories are suitable to connect the weak scale physics 
to the high scale physics in the unified pictures.

Recently, the gauge-Higgs unification~\cite{Manton:1979kb,Hosotani:1983xw,Dvali:2001qr}
in higher-dimensional models is considered
as one of the theoretical origins of Higgs bosons.
Since the extra-dimensional components of gauge fields transform
as scalar fields in four dimension (4D), the zero modes of the extra components 
can be identified as Higgs fields.
The Higgs fields, which break the electroweak gauge symmetry, 
are the doublets under $SU(2)_L$
while the gauge fields are in the adjoint representation,
and thus one needs to extend the electroweak gauge group
in order to realize the gauge-Higgs unification.
The gauge symmetry is broken by the orbifold boundary 
condition~\cite{Kawamura:1999nj,Mimura:2002te},
and the remaining electroweak gauge symmetry is broken further by the Higgs multiplets
which are the zero modes for 
 the broken generators of the bulk gauge group. This is 
the well-known Hosotani mechanism~\cite{Hosotani:1983xw}.
Therefore, the gauge-Higgs unification is compatible with the gauge symmetry unification
in orbifold GUTs~\cite{Hall:2001pg}.
Furthermore, if one realizes the gauge-Higgs unification,
 the Yukawa couplings may arise from the gauge interaction
since the covariant derivative
in higher-dimensional kinetic term, $\bar \Psi \Gamma^M A_M \Psi$, includes Yukawa couplings.
The realistic models in gauge-Yukawa unification are constructed
in 5-dimensional (5D) supersymmetric models~\cite{Burdman:2002se,Gogoladze:2003bb}.
In 6-dimensional (6D) models, the quark and lepton fields can also be unified 
with the gauge multiplet in higher dimensions~\cite{Gogoladze:2003ci,Gogoladze:2003cp}.
In fact, the unification of gauge and Yukawa couplings 
for quarks and leptons in third generation are in very good agreement 
with experiments~\cite{Gogoladze:2003pp},
which gives us a beautiful unified picture at compactification scale.

Although the supersymmetric model can explain the gauge hierarchy problem naturally,
 the supersymmetric Higgs mass $\mu$ need not be hierarchically small compared to
the Planck scale or string scale because it is allowed by gauge symmetry and 
supersymmetry. To break the electroweak gauge symmetry,
the $\mu$-term (and the $B$-term, which is a supersymmetry
breaking Higgs bilinear soft mass term)
must be close to the weak scale and the supersymmetry breaking scale.
Thus, how to explain the weak scale $\mu$-term is an interesting
question. This is known as the $\mu$-problem.
One usually assumes a symmetry to suppress the $\mu$-term~\cite{Lazarides:1998iq}.
For example, the Giudice-Masiero mechanism~\cite{Giudice:1988yz} can 
solve the $\mu$-problem
in the supergravity mediated supersymmetry breaking scenario because
the  $\mu$-term is forbidden by the $R$ symmetry and is generated after the
supersymmetry breaking.
An alternative solution to the $\mu$-problem is 
the Next to the MSSM~\cite{Fayet:1974pd} in which 
a singlet field $S$ and the superpotential $\lambda S H_u H_d$ are introduced.
The explicit $\mu$-term is forbidden by the discrete $Z_3$ symmetry,
and the effective $\mu$-term $\mu = \lambda \langle S \rangle$ is generated after 
$S$ obtains a VEV.
Thus, $\mu$ is naturally around the supersymmetry breaking scale
due to the minimization condition of the scalar potential.
This type of model is easily realized in the context of supersymmetric gauge-Higgs unification
when the bulk gauge symmetry is appropriately extended and the singlet
can be obtained as a massless mode from the bulk gauge multiplet~\cite{Gogoladze:2005az}.
The bulk gauge multiplet can naturally include 
two extra singlet fields $S_1$ and $S_2$,
and the $\kappa S S_1 S_2$ coupling can come from the bulk interaction 
in addition to the coupling $\lambda S H_u H_d$.
Thus, the $\lambda$ and $\kappa$ couplings can be unified with the electroweak gauge couplings 
at the compactification scale.

In this paper, we study various scenarios of coupling unification
in the gauge-Higgs unification, and gauge-Higgs-matter unification where
in addition to the Higgs trilinear couplings $\lambda$ and $\kappa$,
we also have the unification of the Yukawa couplings for quarks and leptons 
in the third generation.
The gauge, Higgs and Yukawa couplings can be unified at compactification scale.
Depending on the bulk gauge group and the hypercharge assignments,
the up-type (top quark and tau-neutrino Dirac) Yukawa couplings or the
down-type (bottom quark and tau) Yukawa couplings
are unified with the gauge couplings.
Especially, the unification of the gauge, Higgs trilinear
 and top Yukawa couplings is an interesting possibility
since it can explain why the top quark is the heaviest fermion
(for the small $\tan\beta$ case).
We also study the numerical predictions of the couplings at low energy.
We will show that the prediction of the Yukawa coupling unification
is in good agreement with experiments,
in particular, the top quark mass prediction is an important result of the 
gauge-Higgs-Yukawa 
coupling unification at the compactification scale.

This paper is organized as follows: 
In Section 2, we study what bulk gauge symmetries realizing 
the gauge-Higgs unification with gauge and Yukawa coupling unification.
The NMSSM superpotential can be generated from the bulk 
gauge interaction when the appropriate bulk gauge symmetries are considered.
In Sections 3 and 4, we will construct 6D $SU(8)$ and $SU(9)$
supersymmetric models on $T^2/Z_6$ orbifold.
In Section 5, we propose 7-dimensional
(7D) $SU(9)$ models on $T^2/Z_6 \times S^1/Z_2$ orbifold.
In Section 6, we present the numerical results of the gauge, Yukawa and/or 
Higgs trilinear coupling unifications.
Section 7 contains our discussions and conclusions.
In Appendix A, we study the
gauge-Higgs unification in 6D supersymmetric models on $T^2/Z_6$ orbifold.


\section{Choice of Bulk Gauge Symmetry}
\setcounter{equation}{0}

The ${\cal N}=1$ supersymmetry in 5D corresponds to ${\cal N}=2$ supersymmetry in 4D,
and the Higgs fields can be contained in ${\cal N}=2$ gauge 
multiplet~\cite{Dvali:2001qr,Burdman:2002se}.
Matter fields are contained in the hypermultiplets.
On the other hand,
${\cal N}=(1,1)$ supersymmetry in 6D corresponds to ${\cal N}=4$ supersymmetry in 4D, 
and thus the models are restricted since
only the gauge multiplet can be introduced in the bulk.
In terms of 4D ${\cal N}=1$ supersymmetry language, the 6D gauge multiplet
contains vector multiplet $V(A_{\mu}, \lambda)$ and three chiral multiplet, 
$\Sigma_i$ $(i=1,2,3)$, 
in adjoint representation of gauge group.
In a sense, the 5D hypermultiplets in the adjoint representation 
can also belong to gauge multiplet, and thus one can consider that the
matter representation  also belong to the bulk gauge multiplets~\cite{Gogoladze:2003ci}.

In this Section, we will study what bulk gauge symmetry can realize the gauge-Higgs unification
with various coupling unifications.

\subsection{Gauge-Higgs Unification and Yukawa Couplings}

The minimal choice of bulk gauge symmetry to realize the gauge-Higgs unification is 
$SU(3)_W$~\cite{Burdman:2002se,Dimopoulos:2002mv}.
The bulk symmetry is broken as $SU(3)_W \rightarrow SU(2)_L \times U(1)_Y$,
and the Higgs doublets ($H_u$ and $H_d$) are included in the gauge multiplet,
${\bf 8} = {\bf 3}_0 + {\bf 2}_{1/2} + {\bf 2}_{-1/2} + {\bf 1}_0\,$.
The hypercharge generator is $T_Y = {\rm diag}\,(1,1,-2)/6$, and 
thus the predicted weak mixing angle
at compactification scale is $\sin^2 \theta_W = 3/4$,
which is too large for usual minimal unification pictures through RGEs.
One needs to add extra fields to change the RGE evolution or 
to consider that the brane localized gauge couplings give
the dominant contributions to the 4D gauge couplings
rather than the bulk gauge coupling.

In order to realize the proper weak mixing angle in the 
minimal unification scenario, $\sin^2 \theta_W =3/8$,
we should consider $SU(4)_W$ bulk gauge symmetry rather than $SU(3)_W$.
The $SU(4)_W$ adjoint representation, $\bf 15$, is decomposed as
${\bf 15} = {\bf (3,1)}_0 + {\bf (1,3)}_{0} + {\bf (2,2)}_{2} + {\bf (2,2)}_{-2} + {\bf (1,1)}_0$
under $SU(4)_W \rightarrow SU(2)_L \times SU(2)_R \times U(1)^\prime$,
and it is easy to see that the Higgs bidoublets are included in the bulk gauge multiplet.
Thus, considering $SU(3)_c \times U(1)_{B-L} \times SU(4)_W$ bulk symmetry,
we can obtain the proper weak mixing angle~\cite{Gogoladze:2003bb,Shafi:2002ck}.
For this, the left-right symmetric base (or Pati-Salam base~\cite{Pati:1974yy}) is useful
because the Higgs bidoublets are not charged under the $U(1)_{B-L}$.
One can also realize the same quantum number 
assignment in $SO(5)_W (\simeq Sp(4)_W)$ as a minimal choice:
$SO(5)$ adjoint ${\bf 10} = ({\bf 3,1})+({\bf 1,3})+({\bf 2,2})$ under 
$SO(5)_W \rightarrow SU(2)_L\times SU(2)_R$.
Of course, in such product  bulk gauge symmetries,
the $B-L$ charge normalization is arbitrary
and thus the weak mixing angle is not completely determined.
One can unify the bulk gauge groups into simple groups, $SO(12)$ 
($SO(11)$) or $SU(8)$.
One can also consider $E_6$ as a GUT group 
using $E_6 \rightarrow SU(5)\times U(1)^2$ branch~\cite{Burdman:2002se,Gogoladze:2003cp}.
Those gauge-Higgs unification can be realized in 5D supersymmetric 
models on $S^1/Z_2$ orbifold.

Interestingly, the bulk vector multiplets for such unification groups 
can also include the matter representations,
and thus the matters can also be unified with the gauge
multiplet in higher-dimensional models~\cite{Gogoladze:2003ci,Gogoladze:2003cp}.
For example, the $SU(8)$ adjoint $\bf 63$ includes
$(\bf 4,2,1)$ and $(\overline{\bf 4},{\bf 1,2})$ as matter representations
in addition to the Higgs bidoublet $(\bf 1,2,2)$
under the $SU(4)_c \times SU(2)_L \times SU(2)_R$ decomposition.
As a result, through the trilinear term in the bulk action 
(see Eq.~(\ref{eq:t2z6action-6})),
\begin{equation}
\int d^2 \theta \:{\rm Tr} \,\frac1{kg^2} \,\left(
- \sqrt2 \,\Sigma_1 [\Sigma_2, \Sigma_3]\right) +{\rm H.C.}\,,
\label{mm1}
\end{equation}
the Yukawa couplings for quark and lepton can be unified with the 
gauge couplings~\cite{Gogoladze:2003pp}.
In the $SO(12)$ and $E_6$ models, the top quark Yukawa coupling can be realized
in the flipped $SU(5)$ and $SU(5)$ branches, respectively~\cite{Gogoladze:2003cp}.
When $SO(16)$ bulk symmetry is considered, the 
second generation of the SM fermions can also be contained in 
the bulk gauge multiplet~\cite{Gogoladze:2003ci}.

\subsection{Gauge-Higgs Unification in the NMSSM}

Next let us consider the NMSSM gauge-Higgs unification~\cite{Gogoladze:2005az}.
The purpose is to include the coupling $\lambda S H_u H_d$ as a zero mode interaction
from the bulk action in Eq.~(\ref{mm1}).
The $SU(3)_W$ adjoint includes one singlet ${\bf 1}_0$ as we have seen,
and the adjoint trilinear interaction ${\bf 8}^3$ includes 
${\bf 2}_{1/2} {\bf 2}_{-1/2} {\bf 1}_0\,$.
However, if the ${\bf1}_0$ component arises from the
zero mode of the chiral multiplets $\Sigma_i$, 
the ${\cal N}=2$ supersymmetry remains in 4D
and it is not a proper situation for our purpose.
The next candidate is $SU(4)_W$.
The $SU(4)_W$ symmetry is broken down to $SU(2)_L \times U(1)_Y \times U(1)^\prime$
and the adjoint representation is decomposed as
\begin{equation}
{\bf15} = \left(
\begin{array}{ccc}
{\bf3}_{0,0} & {\bf2}_{\frac12,-\frac12} & {\bf2}_{\frac12,\frac12}  \\
{\bf2}_{-\frac12,\frac12} & {\bf1}_{0,0} & {\bf1}_{0,1}  \\
{\bf2}_{-\frac12,-\frac12}   & {\bf1}_{0,-1} & {\bf1}_{0,0}  
\end{array}
\right).
\end{equation}
The two subscripts in each decomposed element 
stand for the hypercharges and $U(1)^\prime$ charges, respectively.
The generators for the $U(1)_Y$ and $U(1)^\prime$ are
$T_Y = {\rm diag}\, (1,1,-1,-1)/4$ and $T^\prime = {\rm diag}\, (0,0,1,-1)/2$,
respectively.
In this case we can have $\lambda H_u H_d S$ interaction
 as part of the bulk trilinear gauge interaction. 
The SM singlet field $S$ has $U(1)^\prime$ charge.
At the compactification scale we can have $g_2=\lambda$ where $g_2$ is 
$SU(2)_L$ gauge coupling.  
%
However, the weak mixing angle at the compactification scale 
is predicted to be $\sin^2\theta_W = 2/3$,
which is too large to consider the RGE evolution with the MSSM particle content.
Actually, the hypercharge assignment is incompatible with left-right symmetric basis 
as we have seen previously 
while the $U(1)^\prime$ was the proper charge assignment to 
obtain the left-right symmetric basis.

Employing $SU(5)_W$ bulk symmetry,
we can obtain the NMSSM gauge-Higgs unification with proper hypercharge normalization.
The $SU(5)$ adjoint representation $\bf 24$ is decomposed under 
$SU(2)_L \times U(1)_Y \times U(1)_1 \times U(1)_2$ gauge symmetry as
\begin{equation}
{\bf 24} = \left(
\begin{array}{cccc}
{\bf3}_{Q00} & {\bf2}_{Q12} & {\bf2}_{Q13} & {\bf2}_{Q14} \\
{\bf2}_{Q21} & {\bf1}_{Q00} & {\bf1}_{Q23} & {\bf1}_{Q24} \\
{\bf2}_{Q31} & {\bf1}_{Q32} & {\bf1}_{Q00} & {\bf1}_{Q34} \\
{\bf2}_{Q41} & {\bf1}_{Q42} & {\bf1}_{Q43} & {\bf1}_{Q00} 
\end{array}
\right)\,, \label{SU5-adjoint}
\end{equation}
where the diagonal entries, 1 triplet and 3 singlets corresponds to the unbroken
generators of $SU(5)$.
The subscripts $Qij$, which are anti-symmetric 
($Qij=-Qji$) and $Qii\equiv Q00 =0$, stand for the $U(1)$ charges.
One can calculate $Qij$ by the definition
\begin{equation}
T\cdot{\bf 24} - {\bf 24}\cdot T = Q \cdot {\bf 24},
\end{equation}
where $T$ is the corresponding $U(1)$ generator.
We define the hypercharge generator as 
$T_Y = {\rm diag} \, (1/2, 1/2, 0, 0, -1) + \alpha (1,1,1,1,-4)$.
%
%
%
And the adjoint trilinear coupling includes $S H_u H_d$ as
\begin{equation}
{\rm Tr}\,{\bf 24}^3 \supset {\bf2}_{Q12} {\bf1}_{Q23} {\bf2}_{Q31} \,.
\end{equation}
We identify $\Sigma_1({{\bf1}_{Q23}})$, 
$\Sigma_3({\bf2}_{Q12})$ and $\Sigma_2({\bf2}_{Q31})$
as $S$, $H_u$ and $H_d$, respectively.
The weak mixing angle at the compactification scale is given as
$\sin^2 \theta_W = 1/(4+20 \alpha +40 \alpha^2)$,
where $\alpha$ is a free parameter in the model.
The value of the weak mixing angle is not predicted in this model, 
but it can be consistent with its measured value.

Similar model can be constructed by using $SO(7)_W$, whose dimension is less then $SU(5)_W$. 
Under the decomposition $SO(7) \rightarrow SO(5) \times U(1)^\prime$,
${\bf 21} = {\bf 10}_0 + {\bf 5}_1 + {\bf 5}_{-1} + {\bf 1}_0$,
and under $SO(5) \rightarrow SU(2)_L \times SU(2)_R$,
${\bf 10} = ({\bf 3,1})+({\bf 1,3})+({\bf 2,2})$,
${\bf 5} = ({\bf 2,2})+({\bf 1,1})$.
Thus the adjoint trilinear term
includes $S H_u H_d$, 
${\bf 21}^3 \supset {\bf 10}_0 {\bf 5}_1 {\bf 5}_{-1} 
\supset ({\bf 2,2})_0({\bf 2,2})_1({\bf 1,1})_{-1} \supset H_u H_d S\,$.
In this case, the left-right symmetry is simply embedded, and thus
the weak mixing angle can be $\sin^2 \theta_W = 3/8$ more naturally.

One can see from Eq.~(\ref{SU5-adjoint}) that the $SU(5)_W$ adjoint includes $SU(2)_L$ singlets
and thus the adjoint trilinear term can include the singlet trilinear coupling 
$\kappa S S_1 S_2$ as a zero mode interaction.
In this case, the NMSSM superpotential,
\begin{equation}
W_{\rm NMSSM} = \lambda S H_u H_d - \kappa S S_1 S_2\,,
\label{NMSSM-superpotential}
\end{equation}
can be included in the bulk gauge interaction,
and thus both $\lambda$ and $\kappa$ can be unified with gauge coupling $g_2$
at the compactification scale.
Since the $S_1$ and $S_2$ is needed to be singlets under the SM,
$Q23$, $Q34$ and $Q42$ in Eq.~(\ref{SU5-adjoint}) are all zero for hypercharge and thus
the hypercharge generator is determined as
$T_Y = {\rm diag}\, (3/2,3/2,-1,-1,-1)/5$.
Then the weak mixing angle is calculated as $\sin^2 \theta_W = 5/8$,
and thus extra $U(1)$ symmetry, which mixes with hypercharge,
 is needed to realize the proper weak
mixing angle in the usual minimal unification scenario.
When we consider the $SU(6)_W$, the NMSSM superpotential in Eq.~(\ref{NMSSM-superpotential})
is included in the bulk interaction in compatible with proper weak mixing angle,
since $SU(6)_W$ includes the trinification basis, 
$SU(6)_W \rightarrow SU(3)_L \times SU(3)_R \times U(1)^\prime$ :
${\bf 35} = ({\bf 8,1})+({\bf 1,8})+({\bf 3,\bar 3})+({\bf \bar 3,3})+ ({\bf 1,1})$,
and ${\bf 35}^3 \supset ({\bf 1,8})({\bf 3, \bar 3})({\bf \bar 3,3})$ can include
both $\lambda$ and $\kappa$ couplings.
One can also use the branch $(F_4)_W \rightarrow SU(3)_L \times SU(3)_R$
to make the NMSSM superpotential.
In such trinification assignment, the $B-L$ charge is also 
embedded in the $SU(6)_W$ and $(F_4)_W$,
and the weak mixing angle is predicted properly.
The $SO(8)_W$ bulk interaction which includes the NMSSM superpotential is also compatible with
proper weak mixing angle, although it is not completely determined 
since the $B-L$ charge does not belong to $SO(8)_W$.

We note that 
there are five SM neutral complex scalar fields
and three phase symmetries in the renormalizable superpotential 
in Eq.~(\ref{NMSSM-superpotential}). 
One of these is related to the $U(1)_Y$,
and the other two combinations are unwanted global symmetries,
which implies that there are two massless Goldstone bosons 
after electroweak symmetry breaking.
To avoid the problem, 
we assume that there are non-renormalizable couplings involving 
$S$, $S_1$ and $S_2$, 
such that the couplings break extra $U(1)$ symmetries having in the superpotential.
Giving a suitable choice of charge assignment for the singlet fields,
we can generate non-renormalizable couplings or tiny mass term
for $S$, $S_1$ and $S_2$ fields.
In that way, we can make the model not having the massless Goldstone bosons in the spectrum,
and the model does not have
neither domain wall problem~\cite{Abel:1995wk} nor tadpole corrections.

We also note that the renormalizable superpotential, Eq.~(\ref{NMSSM-superpotential}),
is different from the one in the Next to the MSSM \cite{Fayet:1974pd}.
The singlet trilinear coupling is not a self-cubic coupling contrary to the Next to the MSSM.
However, as long as the orbifold model is concerned,
it is more natural to have the coupling, $\kappa S S_1 S_2$,
if the renormalizable coupling needed in the Higgs superpotential
is included in the bulk interaction.
As we have seen the construction, the $\kappa$ coupling is not necessarily unified 
in the bulk interaction depending on the bulk gauge symmetry.
If the singlet trilinear coupling is not in the bulk interaction,
one can assume that the brane-localized interactions include 
the $S^3$ term having in the Next to the MSSM.
In any case, it is flexible to construct such types of Non-Minimal Supersymmetric Standard Models.

In the same way as before, when the grand unified bulk gauge symmetry is considered,
the Yukawa couplings for the SM fermions can also be unified with the gauge couplings
in addition to the NMSSM couplings.
One can consider the $SU(8)$ and $SU(9)$ bulk gauge symmetry as grand unified groups
(instead of $SU(5)_W$ and $SU(6)_W$ for the electroweak sector).
The $SU(8)$ bulk gauge interaction includes Eq.~(\ref{NMSSM-superpotential})
with grand unification in the branch 
$SU(8) \rightarrow SU(5)_{\rm GUT} \times SU(3) \times U(1)^\prime$
(but Yukawa couplings for fermions are not unified in this case).
When Pati-Salam branch of $SU(8)$ is considered, some of the  Yukawa coupling for fermions and
 the $\lambda$ coupling in the NMSSM can be unified with the gauge couplings.
In the case of $SU(9)$,  the gauge couplings, Higgs trilinear couplings
and the up-type or down-type Yukawa couplings for the SM fermions
can be unified.

 We will construct the concrete models  in the next three Sections, realizing
 various possibilities.

\section{The $SU(8)$ Models}
\setcounter{equation}{0}

To break the $SU(8)$ gauge symmetry, we choose
the following $8\times 8$ matrix representation for 
$R$,
\begin{equation}
R = {\rm diag}\, \left(\omega^{n_1}, \omega^{n_1}, \omega^{n_1},
 \omega^{n_2}, \omega^{n_2}, \omega^{n_3}, \omega^{n_4}, \omega^{n_5} \right)\,,
\end{equation}
where $n_i \neq n_j$ $(i \neq j)$.
Then, $SU(8)$ is broken as 
\begin{eqnarray}
 SU(8) \rightarrow SU(3)_c\times SU(2)_L\times U(1)_Y \times U(1)_1
\times U(1)_2 \times U(1)_3 \,.
\end{eqnarray}
Without loss of generality, we can take $n_1 =0$.

We define the generators for $U(1)_Y \times U(1)_1
\times U(1)_2 \times U(1)_3$ as following
\begin{eqnarray}
T_Y &\!\!=&\!\!  {\rm diag}\left( -{1\over 3}, 
-{1\over 3}, -{1\over 3}, {1\over 2}, {1\over 2}, 
0, 0, 0 \right), 
\label{GU1Y} \\
T_1 &\!\!=&\!\! {\rm diag}\left( {1\over 5},
{1\over 5}, {1\over 5}, {1\over 5}, {1\over 5},
-{1\over 3}, -{1\over 3}, -{1\over 3} \right), 
\label{GU1A} \\
T_2 &\!\!=&\!\!  {\rm diag}\left(
0, 0, 0, 0, 0, {1\over 2}, - {1\over 2}, 0 \right), 
\label{GU1B} \\
T_3 &\!\!=&\!\! {\rm diag}\left(
0, 0, 0, 0, 0, {1\over 2}, {1\over 2}, -1 \right).
\label{GU1C}
\end{eqnarray}
 
The $SU(8)$ adjoint representation $\mathbf{63}$
is decomposed under the 
$SU(3)_c\times SU(2)_L\times U(1)_Y \times U(1)_1
\times U(1)_2 \times U(1)_3$ gauge symmetry as
\begin{equation}
\mathbf{63} = \left(
\begin{array}{ccccc}
\mathbf{(8,1)}_{Q00} & \mathbf{(3, \bar 2)}_{Q12} 
& \mathbf{(3, 1)}_{Q13} & \mathbf{(3,1)}_{Q14}
 & \mathbf{(3,1)}_{Q15} \\
 \mathbf{(\bar 3,  2)}_{Q21} & \mathbf{(1,3)}_{Q00}
& \mathbf{(1, 2)}_{Q23} & \mathbf{(1, 2)}_{Q24}
& \mathbf{(1, 2)}_{Q25} \\
\mathbf{(\bar 3, 1)}_{Q31} & \mathbf{(1, \bar 2)}_{Q32}
& \mathbf{(1, 1)}_{Q00} & \mathbf{(1, 1)}_{Q34}
& \mathbf{(1, 1)}_{Q35} \\
\mathbf{(\bar 3, 1)}_{Q41} & \mathbf{(1, \bar 2)}_{Q42}
& \mathbf{(1, 1)}_{Q43} & \mathbf{(1, 1)}_{Q00}
& \mathbf{(1, 1)}_{Q45} \\
\mathbf{(\bar 3, 1)}_{Q51} & \mathbf{(1, \bar 2)}_{Q52}
& \mathbf{(1, 1)}_{Q53} & \mathbf{(1, 1)}_{Q54}
& \mathbf{(1, 1)}_{Q00} 
\end{array}
\right) +  \mathbf{(1,1)}_{Q00} \,,
\end{equation}
where the  $\mathbf{(1,1)}_{Q00}$ in the third, fourth and fifth
diagonal entries of the matrix, and the last term $\mathbf{(1,1)}_{Q00}$ 
denote the gauge fields for the $U(1)_Y \times U(1)_1
\times U(1)_2 \times U(1)_3$ gauge symmetry.
The subscripts $Qij$, which are anti-symmetric
($Qij=-Qji$), are the charges under
the $U(1)_Y \times U(1)_1
\times U(1)_2 \times U(1)_3$ gauge symmetry, which can be easily calculated
by the respective definitions of the $U(1)$ generators,
\begin{eqnarray}
&& Q00= ({0}, {0}, {0}, {0}) \,,~~
 Q12= ({-{5\over 6}}, {0}, {0}, {0}) \,,~~
Q13=({-{1\over 3}}, {{8\over{15}}}, 
{-{1\over 2}}, {-{1\over 2}}) \,, \\
&& Q14=({-{1\over 3}}, {{8\over{15}}}, 
{{1\over 2}}, {-{1\over 2}}) \,,~~
 Q15=({-{1\over 3}}, {{8\over{15}}}, 
{0}, {1}) \,,~~
 Q23=({{1\over 2}}, {{8\over{15}}}, 
{-{1\over 2}}, {-{1\over 2}}) \,, \\
&& Q24=({{1\over 2}}, {{8\over{15}}}, 
{{1\over 2}}, {-{1\over 2}}) \,,~~
 Q25=({{1\over 2}}, {{8\over{15}}}, 
{0}, {1})\,,~~
Q34=({0}, {0}, {1}, {0}) \,, \\
&&  Q35=({0}, {0}, {{1\over 2}}, 
{{3\over 2}})\,,~~
 Q45=({0}, {0}, {-{1\over 2}}, 
{{3\over 2}}) \,.
\end{eqnarray}

The $Z_6$ transformation property for the decomposed components
of $V$, $\Sigma_1$, $\Sigma_2$,  and $\Sigma_3$ are given in the notation in Appendix A,
\begin{equation}
V^{(ij)} : \omega^{n_i-n_j}\,, \quad 
\Sigma_1^{(ij)} : \omega^{n_i-n_j-1}\,,\quad 
\Sigma_2^{(ij)} : \omega^{n_i-n_j -1-m} \,,\quad 
\Sigma_3^{(ij)} : \omega^{n_i-n_j +2+m} \,.
\label{Orbifold-conditions}
\end{equation}

We can have several models which are quite similar. So, for simplicity, 
we only present one model which needs less 3-brane localized
exotic quarks. These 3-brane localized exotic quarks and some extra 
particles from the zero modes of the chiral multiplets $\Sigma_i$
are vector-like under the SM gauge symmetry
and can obtain the vector-like masses after the extra $U(1)_1
\times U(1)_2 \times U(1)_3$ gauge symmetry 
is broken at the GUT scale.

In our model, we choose $m=1$, and
\begin{equation}
n_2=5\,, \ n_3=4\,,\ n_4=2\,, \ n_5=1 \,.
\end{equation}
The corresponding zero modes from  the chiral multiplets
$\Sigma_1$, $\Sigma_2$ and $\Sigma_3$ are given in
the Table \ref{Spectrum0}.

\begin{table}[t]
\caption{The zero modes of the chiral multiplets 
$\Sigma_1$, $\Sigma_2$ and $\Sigma_3$
in the 6D orbifold $SU(8)$ model.
\label{Spectrum0}}
\begin{center}
\begin{tabular}{|c|c@{:}cc@{:}cc@{:}cc@{:}c|}
\hline 
Chiral Fields & \multicolumn{8}{c|}{Zero Modes}  \\
\hline\hline
$\Sigma_1$ & $Q_X\,$ & ~$\mathbf{(3, \bar 2)}_{Q12}\,$; &
$H_u'\,$ & ~$\mathbf{(1, 2)}_{Q23}\,$; &
$S\,$ & ~$\mathbf{(1, 1)}_{Q45}\,$; &
$\overline{D}_{\delta}\,$ & ~$\mathbf{(\bar 3, 1)}_{Q51}$ \\
\hline
$\Sigma_2$ & $D_{\delta}\,$ & ~$\mathbf{(3, 1)}_{Q13}\,$; &
$S_2\,$ & ~$\mathbf{(1, 1)}_{Q34}\,$; &
$\overline{D}_X\,$ & ~$\mathbf{(\bar 3, 1)}_{Q41}\,$; &
$H_d\,$ & ~$\mathbf{(1, \bar 2)}_{Q52}$ \\
\hline
$\Sigma_3$ & $H_u\,$ & ~$\mathbf{(1, 2)}_{Q24}\,$; &
$S_1\,$ & ~$\mathbf{(1, 1)}_{Q53}\,$; &
$S_1'\,$ & ~$\mathbf{(1, 1)}_{Q35}\,$; &
$H_d'\,$ & ~$\mathbf{(1, \bar 2)}_{Q42}$ \\
\hline 
\end{tabular}
\end{center}
\end{table}

{}From the 6D bulk interaction in Eq.~(\ref{mm1}),
we obtain the Yukawa terms
\begin{eqnarray}
{\cal S} &\!\!=&\!\! 
\int d^6 x \left[ \int d^2 \theta \ g_6 \left(
S H_u H_d - S S_1 S_2 - Q_X \overline{D}_X H_u +
S_2 H_u' H_d' 
\right. \right. \nonumber\\ && \left. \left.
\qquad \qquad \qquad \qquad 
- S_1' H_u' H_d 
+ S_1' D_{\delta}
\overline{D}_{\delta} \right)
 + {\rm H. C.}\right],
\label{6D-bulk-SU8}
\end{eqnarray}
where $g_6$ is the 6D bulk gauge coupling.

Because the $U(1)_1
\times U(1)_2 \times U(1)_3$ gauge symmetry
can be broken at the GUT scale, the exotic quarks 
$Q_X$, $\overline{D}_X$, 
$D_{\delta}$, $\overline{D}_{\delta}$, and
the doublets $H_u'$ and
$H_d'$ can become heavy after these 
extra $U(1)$ gauge symmetry breakings.
To achieve this, on the 3-brane at
the $Z_6$ fixed point, for example, $z=0$,
we introduce two exotic quarks $\overline{Q}'_X$
and $D'_{\delta}$ with respectively quantum numbers 
$\mathbf{(\bar 3,  2)}_{{(5/6, 0, -1, 0)}}$ and
$\mathbf{(3, 1)}_{{(-1/3, 8/15, -1, 1)}}$
under the $SU(3)_C\times SU(2)_L\times U(1)_Y \times U(1)_1
\times U(1)_2 \times U(1)_3$ gauge symmetry.
We also introduce a SM singlet Higgs field
${\widetilde S}_2$ which has the same quantum number as that of
$S_2$ and is localized on the 3-brane at $z=0$.
After ${\widetilde S}_2$ gets VEV, the exotic quarks and
 the  doublets $H_u'$ and $H_d'$ can obtain the 
vector-like masses through the following 3-brane localized
superpotential
\begin{equation}
W =  {\widetilde S}_2 H_u' H_d'  
+ {\widetilde S}_2 D_{\delta} \overline{D}_X
+ {\widetilde S}_2 Q_X \overline{Q}'_X
+ {\widetilde S}_2 D'_{\delta} \overline{D}_{\delta} ~.~\,
\end{equation}
Similarly, $S_1'$ can also become heavy, but, we require that
it do not get a VEV due to the superpotential $S_1' H_u' H_d$.

Furthermore, we would like to point out that
 if we consider $\overline{D}_{\delta}$
as the right-handed down-type quark in the supersymmetric SM,
 we do not need to introduce the exotic quark $D'_{\delta}$
on the 3-brane at $z=0$. However, $Q_X$ can not be considered
as the quark doublet because its hypercharge is ${-5/6}$.

After the $U(1)_1
\times U(1)_2 \times U(1)_3$ gauge symmetry
is broken, we can have the relevant superpotential
\begin{equation}
{\cal S} = \int d^6 x \left[ \int d^2 \theta \ g_6 \left(
S H_u H_d - S S_1 S_2  \right)
 + {\rm H. C.}\right] .
\end{equation}
Integrating out the extra dimensions, 
we obtain the NMSSM superpotential in Eq.~(\ref{NMSSM-superpotential}).

For simplicity, we assume that the compactification scale is the GUT scale
and we neglect the brane-localized gauge kinetic terms which can be suppressed by
large volume of the extra dimensions.
The weak mixing angle is calculated to be $\sin^2 \theta_W =3/8$,
as long as we use the definition of hypercharge generator in Eq.~(\ref{GU1Y}).
The $\lambda$ and $\kappa$ couplings in the superpotential can be
unified with the gauge couplings at the compactification scale,
\begin{equation}
g_1 = g_2 = g_3 = \lambda = \kappa = g_6/\sqrt{V}\,,
\end{equation}
where $V$ is the volume of extra dimensions.
The hypercharge gauge coupling is normalized as $g_1 = \sqrt{5/3}\, g_Y$.
However, the hypercharge normalization cannot be determined completely
as long as only Higgs fields are in the bulk.
In fact, $T_Y + \alpha T_1$ can be a hypercharge generator, where $\alpha$ is a free parameter.
In order to determine the hypercharge normalization completely,
quarks and/or leptons are needed to be from the bulk gauge multiplet.
For example, if we identify $\overline{D}_\delta$ as the 
right-handed down-type quark field,
$\alpha$ is determined to be zero, and the charge quantization is fixed. 

We have assumed that the hypercharge generator is $SU(5)$-type unification
as in Eq.~(\ref{GU1Y}).
In this assignment, no Yukawa coupling for quark/lepton 
is included in the bulk interaction.
Suppose that the hypercharge generator is redefined as
\begin{equation}
T_Y = {\rm diag}\, (\frac16,\frac16,\frac16,0,0,\frac12,-\frac12,-\frac12)\,,
\end{equation}
 then, the hypercharges of the zero modes in the chiral multiplets can change.
We list the zero modes in chiral multiplets in Table~\ref{Spectrum0-2} with 
an appropriate notation to see their hypercharges.
\begin{table}[t]
\caption{The zero modes of the chiral multiplets 
$\Sigma_1$, $\Sigma_2$ and $\Sigma_3$
in the 6D orbifold $SU(8)$ model with Pati-Salam charge assignment.
\label{Spectrum0-2}}
\begin{center}
\begin{tabular}{|c|c@{:}cc@{:}cc@{:}cc@{:}c|}
\hline 
Chiral Fields & \multicolumn{8}{c|}{Zero Modes}   \\
\hline\hline
$\Sigma_1$ & $Q\,$ &  ~$\mathbf{(3, \bar 2)}_{Q12}\,$; &
$H_d\,$ & ~$\mathbf{(1, 2)}_{Q23}\,$; &
$\overline{N}\,$ & ~$\mathbf{(1, 1)}_{Q45}\,$; &
$\overline{U}_{\delta}\,$ & ~$\mathbf{(\bar 3, 1)}_{Q51}$ \\
\hline
$\Sigma_2$ & $D_{\delta}\,$ & ~$\mathbf{(3, 1)}_{Q13}\,$; &
$\overline{E}_\delta\,$ & ~$\mathbf{(1, 1)}_{Q34}\,$; &
$\overline{U}\,$ & ~$\mathbf{(\bar 3, 1)}_{Q41}\,$; &
$L\,$ & ~$\mathbf{(1, \bar 2)}_{Q52}$ \\
\hline
$\Sigma_3$ & $H_u\,$&  ~$\mathbf{(1, 2)}_{Q24}\,$; &
$E_\delta\,$ & ~$\mathbf{(1, 1)}_{Q53}\,$; &
$\overline{E}\,$ & ~$\mathbf{(1, 1)}_{Q35}\,$; &
$L^\prime\,$ & ~$\mathbf{(1, \bar 2)}_{Q42}$ \\
\hline 
\end{tabular}
\end{center}
\end{table}
This hypercharge generator can be identified as the Pati-Salam hypercharge assignment
embedded in $SU(8)$.
Then, we find that the 6D bulk interaction in Eq.~(\ref{6D-bulk-SU8}) includes
the top, tau and tau Dirac-neutrino Yukawa couplings,
\begin{eqnarray}
{\cal S} &\!\!=&\!\! 
\int d^6 x \left[ \int d^2 \theta \ g_6 \left(
L \overline{N} H_u - \overline{N} E_\delta \overline{E} - Q \overline{U} H_u +
\overline{E}_\delta H_d L^\prime 
\right. \right. \nonumber\\ && \left. \left.
\qquad \qquad \qquad \qquad 
- L \overline{E} H_d  
+ \overline{E} D_{\delta} \overline{U}_{\delta} \right)
 + {\rm H. C.}\right].
\label{6D-bulk-SU8-2}
\end{eqnarray}
%
The Dirac-neutrino Yukawa coupling can be considered as $\lambda S H_u H_d$ coupling,
but the singlet trilinear term is not included 
since the  $S_1$ and $S_2$ have non-zero hypercharges in this case.
When we extend bulk gauge group, both the 
NMSSM superpotential and the quark/lepton Yukawa couplings
can arise from the bulk gauge interaction. 
We will study the $SU(9)$ bulk gauge symmetry in the next two Sections.

\section{6D $SU(9)$ Models}
\setcounter{equation}{0}

To break the $SU(9)$ gauge symmetry, we choose
the following $9\times 9$ matrix representation for $R$
\begin{equation}
R = {\rm diag} \left(\omega^{n_1}, \omega^{n_1}, \omega^{n_1}, \omega^{n_1},
 \omega^{n_2}, \omega^{n_2}, \omega^{n_3}, \omega^{n_4}, \omega^{n_5} \right)~,~\,
\end{equation}
where $n_i \neq n_j$ $(i\neq j)$. 
Without loss of generality, we choose $n_1 = 0$.
Then, $SU(9)$ is broken as 
\begin{equation}
 SU(9) \rightarrow SU(4)\times SU(2)_L\times U(1)' \times U(1)_{\alpha}
\times U(1)_{\beta} \times U(1)_{\gamma} \,.
\end{equation}

We define the generators for $U(1)' \times U(1)_{\alpha}
\times U(1)_{\beta} \times U(1)_{\gamma}$ as following :
\begin{eqnarray}
T^\prime &\!\!=&\!\! {\rm diag}\left( 0, 0, 0, 0, 0, 0, +1, -1,0 \right) , \\
\label{GU1P}
T_{{\alpha}} &\!\!=&\!\!  
{\rm diag}\left( 0, 0, 0, 0, +1, +1, -1, -1, 0 \right) , \\
\label{GU1A2}
T_{{\beta}} &\!\!=&\!\!  
{\rm diag}\left( +1, +1, +1, +1, -1, -1, -1, -1, 0 \right) , \\
\label{GU1B2}
T_{{\gamma}} &\!\!=&\!\! 
{\rm diag}\left( +1, +1, +1, +1, +1, +1, +1, +1, -8 \right) .
\label{GU1C2}
\end{eqnarray}
 
The $SU(9)$ adjoint representation $\mathbf{80}$
is decomposed under the 
$SU(4)\times SU(2)_L\times U(1)' \times U(1)_{\alpha}
\times U(1)_{\beta} \times U(1)_{\gamma}$ gauge symmetry as
\begin{equation}
\mathbf{80} = \left(
\begin{array}{ccccc}
\mathbf{(15,1)}_{Q00} & \mathbf{(4, \bar 2)}_{Q12} 
& \mathbf{(4, 1)}_{Q13} & \mathbf{(4,1)}_{Q14}
 & \mathbf{(4,1)}_{Q15} \\
 \mathbf{(\bar 4,  2)}_{Q21} & \mathbf{(1,3)}_{Q00}
& \mathbf{(1, 2)}_{Q23} & \mathbf{(1, 2)}_{Q24}
& \mathbf{(1, 2)}_{Q25} \\
\mathbf{(\bar 4, 1)}_{Q31} & \mathbf{(1, \bar 2)}_{Q32}
& \mathbf{(1, 1)}_{Q00} & \mathbf{(1, 1)}_{Q34}
& \mathbf{(1, 1)}_{Q35} \\
\mathbf{(\bar 4, 1)}_{Q41} & \mathbf{(1, \bar 2)}_{Q42}
& \mathbf{(1, 1)}_{Q43} & \mathbf{(1, 1)}_{Q00}
& \mathbf{(1, 1)}_{Q45} \\
\mathbf{(\bar 4, 1)}_{Q51} & \mathbf{(1, \bar 2)}_{Q52}
& \mathbf{(1, 1)}_{Q53} & \mathbf{(1, 1)}_{Q54}
& \mathbf{(1, 1)}_{Q00} 
\end{array}
\right) +  \mathbf{(1,1)}_{Q00} \,,
\end{equation}
where the  $\mathbf{(1,1)}_{Q00}$ in the third, fourth and fifth
diagonal entries of the matrix, and the last term $\mathbf{(1,1)}_{Q00}$ 
denote the gauge fields for the $U(1)' \times U(1)_{\alpha}
\times U(1)_{\beta} \times U(1)_{\gamma}$ gauge symmetry. Moreover, 
the subscripts $Qij$, which are anti-symmetric
($Qij=-Qji$), are the charges under
the $U(1)' \times U(1)_{\alpha}
\times U(1)_{\beta} \times U(1)_{\gamma}$ gauge symmetry
\begin{eqnarray}
&& Q00 = ({0}, {0}, {0}, {0}) \,,\quad 
 Q12=({0}, {-1}, {2}, {0}) \,,\quad 
Q13=({-1}, {1}, {2}, {0}) \,, \\
&& Q14=({1}, {1}, {2}, {0}) \,, \quad
Q15=({0}, {0}, {1}, {9})\,, \quad
Q23=({-1}, {2}, {0}, {0}) \,, \\
&& Q24=({1}, {2}, {0}, {0}) \,, \quad
 Q25=({0}, {1}, {-1}, {9}) \,, \quad
Q34=({2}, {0}, {0}, {0}) \,, \\
&&  Q35=({1}, {-1}, {-1}, {9}) \,, \quad
Q45=({-1}, {-1}, {-1}, {9}) \,.
\end{eqnarray}

The $Z_6$ transformation property for the decomposed components
of $V$, $\Sigma_1$, $\Sigma_2$,  and $\Sigma_3$ are given similarly 
as those in Eq.~(\ref{Orbifold-conditions}).

We can have
 several models which are quite similar. So, for simplicity, 
we only present one model with the up-type partial Yukawa unification, and
one similar model with the down-type partial Yukawa unification.
There are some extra 
particles from the zero modes of the chiral multiplets $\Sigma_i$.
To give them very heavy masses, we introduce 
3-brane localized additional particles.
 These 3-brane localized additional particles and the  extra 
particles from the zero modes of the chiral multiplets $\Sigma_i$
are vector-like under the SM gauge symmetry
and can obtain the vector-like masses after the 
$U(1)'$ gauge symmetry 
is broken at the GUT scale.

\subsection{Up-Type Partial Unification}

We define the generator for $U(1)_{I_{3R}}$ as
\begin{eqnarray}
T_{{I_{3R}}} &\equiv& {1\over 4} T_{{\alpha}}
+ {1\over 8} T_{{\beta}} + {1\over {24}} T_{{\gamma}}
\nonumber\\ 
&=& {\rm diag}\left( +{1\over 6}, +{1\over 6}, +{1\over 6}, 
+{1\over 6}, +{1\over 6}, +{1\over 6}, 
-{1\over 3}, -{1\over 3}, -{1\over 3} \right).\,
\label{U1B-L-U}
\end{eqnarray}
Then, ${\rm Tr} [T_{{I_{3R}}}^2]=1/2$.

We assume that the $SU(4)\times SU(2)_L \times U(1)_{I_{3R}}$
gauge symmetry is broken down to the SM gauge symmetry
at the compactification scale by the Higgs fields with
quantum numbers $(\overline{\bf 4}, {\bf 1}, -1/2)$ and
$( {\bf 4}, {\bf 1}, 1/2)$ under the $SU(4)\times SU(2)_L \times U(1)_{I_{3R}}$
gauge symmetry, {\it i.e.},
 the same quantum numbers as those of the right-handed
neutrino and its Hermitian conjugate.
Then, we obtain that $\sin^2 \theta_W = 3/8$ at the unification scale.
The $U(1)_{B-L}$ generator is embedded in $SU(4)$ as
\begin{equation}
T_{B-L} = {\rm diag}  \left( \frac13, \frac13,\frac13, -1, 0,0,0,0,0 \right) .
\end{equation}
The hypercharge generator is
$T_Y = T_{I_{3R}} + \frac12 T_{B-L}\,$, and the normalization is determined 
as $g_Y = \sqrt{3/5} \, g_1$.


For the transformation, Eq. (3.12) of the vector and the chiral
multiplets, we choose $m=0$ and 
\begin{equation}
n_2=5 \,,~~ n_3=3 \,,~~ n_4=4 \,,~~ n_5=2 \,.
\label{D6-ni}
\end{equation}
The zero modes from  the chiral multiplets
$\Sigma_1$, $\Sigma_2$ and $\Sigma_3$ are given in
the Table \ref{Spectrum-UP-6D}.
It is easy to check that for $L_2$, $L_3$, $S$, $S'$,
$S_1$, $S'_1$ and $S_2$, the $U(1)_{I_{3R}}$ charges are
zero; for $H_u$, $H'_u$ and $L_X$, the $U(1)_{I_{3R}}$ charges are
$1/2$; and for $H_d$ and $R_3^u$,
 the $U(1)_{I_{3R}}$ charges are $-1/2$.

\begin{table}[t]
\caption{The zero modes of the chiral multiplets 
$\Sigma_1$, $\Sigma_2$ and $\Sigma_3$
in the 6D orbifold $SU(9)$ model
with up-type partial unification.
\label{Spectrum-UP-6D}}
\begin{center}
\begin{tabular}{|c|c@{:}cc@{:}cc@{:}cc@{:}c|}
\hline 
Chiral Fields & \multicolumn{8}{c|}{Zero Modes}  \\
\hline\hline
$\Sigma_1$ & $L_3\,$ &~$\mathbf{(4, \bar 2)}_{Q12}\,$; &
$H_u'\,$ & ~$\mathbf{(1, 2)}_{Q24}\,$; & 
$S'_1\,$&~$\mathbf{(1, 1)}_{Q35}\,$; &
$S\,$&~$\mathbf{(1, 1)}_{Q43}$ \\
\hline
$\Sigma_2\,$ & $L_2\,$ &~$\mathbf{(4, \bar 2)}_{Q12}\,$; &
$H_u\,$& ~$\mathbf{(1, 2)}_{Q24}\,$; &
$S_1\,$& ~$\mathbf{(1, 1)}_{Q35}\,$; &
$S'\,$ & ~$\mathbf{(1, 1)}_{Q43}$ \\
\hline
$\Sigma_3$ & $L_X\,$ & ~$\mathbf{(4, 1)}_{Q15}\,$; &
$H_d\,$ & ~$\mathbf{(1, \bar 2)}_{Q32}\,$; &
$R_3^u\,$ & ~$\mathbf{(\bar 4, 1)}_{Q41}\,$; &
$S_2\,$ & ~$\mathbf{(1, 1)}_{Q54}$ \\
\hline 
\end{tabular}
\end{center}
\end{table}

{}From the trilinear term in the 6D bulk action in Eq.~(\ref{mm1}),
we obtain the Yukawa terms
\begin{eqnarray}
{\cal S} &\!\!=&\!\! \int d^6 x \left[ \int d^2 \theta \ g_6 \left(
L_3 R_3^u H_u + L_2 R_3^u H'_u +
S H_d H_u + S' H_d H'_u \right.\right.\nonumber \\
&& \qquad \qquad \qquad \qquad 
+ S S_1 S_2 +
S' S_1' S_2 ) 
 + {\rm H. C.}\biggr].
\end{eqnarray}

The $L_2$ can be considered as the left-handed fermions in
the second family in the SM.
Because the $U(1)'$ gauge symmetry
can be broken at the GUT scale, the extra particles
$H_u'$, $L_X$, $S'$ and $S'_1$ can become very heavy after 
it is broken. In addition, although we can not distinguish
the fields ($H_u$, $H'_u$), ($S$, $S'$) and
($S_1$, $S'_1$) by gauge symmetry, we can distinguish
them via the $R$ symmetry~\cite{Li:2003ee}. 
The $R$ symmetry for the
 $T^2/Z_6$ orbifold is 
$SO(2)_{56} \times U(1)_{4_+} \times U(1)_{4_-}$.
Under this $R$ symmetry, the quantum numbers for
$z$, $\theta$, $V$, $\Sigma_1$, $\Sigma_2$,
and $\Sigma_3$ are ${(1, 0, 0)}$, 
${(-1/2, -1/2, 0)}$, ${(0, 0, 0)}$,
${(-1, 0, 0)}$, ${(0, -1/2, -1/2)}$,
and ${(0, -1/2, 1/2)}$, 
respectively (For details, see Ref.~\cite{Li:2003ee}.).
To give the masses to $H_u'$ and $L_X$, 
on the 3-brane at
the $Z_6$ fixed point, for example, $z=0$,
we introduce three 3-brane localized fields 
${\widetilde S}$, ${\widetilde H}_d$
and ${\widetilde R}_X$ with 
respectively gauge quantum numbers 
$\mathbf{(1,  1)}_{{(-1, 0, 0, 0)}}$,
$\mathbf{(1,  2)}_{{(0, -2, 0, 0)}}$
and $\mathbf{(\bar 4,  1)}_{{(1, 0, -1, -9)}}$
under the $SU(4) \times SU(2)_L\times U(1)' \times U(1)_{\alpha}
\times U(1)_{\beta} \times U(1)_{\gamma}$ gauge symmetry.
We also assume that 
under the $R$ symmetry, the quantum numbers for
${\widetilde S}$, ${\widetilde H}_d$ and ${\widetilde R}_X$
are ${(0, -1/2, -1/2)}$, ${(0, -1/2, 1/2)}$, and 
${(-1, 0, 0)}$, respectively.
After ${\widetilde S}$ gets VEV, the ($H_u'$, ${\widetilde H}_d$) and
($L_X$, ${\widetilde R}_X$)  can obtain the 
vector-like masses through the following 3-brane localized
superpotential
\begin{equation}
W = {\widetilde S} H_u' {\widetilde H}_d
+ {\widetilde S} L_X {\widetilde R}_X \,.
\end{equation}
However, the term ${\widetilde S} H_u {\widetilde H}_d $
is forbidden by the $R$ symmetry.
Similarly, $S'$ and $S_1'$ can also become very heavy.

In short, after the $U(1)'$ gauge symmetry is broken,
we can have the relevant superpotential
\begin{eqnarray}
{\cal S} &=& \int d^6 x \left[ \int d^2 \theta \ g_6 \left(
L_3 R_3^u H_u  + S H_d H_u  + S S_1 S_2  \right)
 + {\rm H. C.}\right].
\end{eqnarray}
 giving rise to the unification of the gauge couplings with
 $\lambda$, $\kappa$ and $y_t$.

\subsection{Down-Type Partial Unification}

We define the generator for $U(1)_{I_{3R}}$ as
\begin{eqnarray}
T_{{I_{3R}}} &\equiv& -{1\over 4} T_{{\alpha}}
- {1\over 8} T_{{\beta}} - {1\over {24}} T_{{\gamma}}
\nonumber\\ 
&=& {\rm diag}\left( -{1\over 6}, -{1\over 6}, -{1\over 6}, 
-{1\over 6}, -{1\over 6}, -{1\over 6}, 
+{1\over 3}, +{1\over 3}, +{1\over 3} \right) .
\label{U1B-L-D}
\end{eqnarray}

Similar to the above up-type partial unification model, we
choose $m=0$ and $n_i$ in Eq.~(\ref{D6-ni}).
The zero modes from  the chiral multiplets
$\Sigma_1$, $\Sigma_2$ and $\Sigma_3$ are given in
the Table \ref{Spectrum-DOWN-6D}.
It is easy to check that for $L_2$, $L_3$, $S$, $S'$,
$S_1$, $S'_1$ and $S_2$, the $U(1)_{I_{3R}}$ charges are
zero; for $H_u$ and $R_3^d$,
 the $U(1)_{I_{3R}}$ charges are $1/2$;
and for $H_d$, $H'_d$ and $L_X$, the $U(1)_{I_{3R}}$ charges are
$-1/2$.

\begin{table}[t]
\caption{The zero modes of the chiral multiplets 
$\Sigma_1$, $\Sigma_2$ and $\Sigma_3$
in the 6D orbifold $SU(9)$ model
with down-type partial unification.
\label{Spectrum-DOWN-6D}}
\begin{center}
\begin{tabular}{|c|c@{:}cc@{:}cc@{:}cc@{:}c|}
\hline 
Chiral Fields & \multicolumn{8}{c|}{Zero Modes}  \\
\hline\hline
$\Sigma_1$ & $L_3\,$ &~$\mathbf{(4, \bar 2)}_{Q12}\,$; &
$H_d'\,$ & ~$\mathbf{(1, 2)}_{Q24}\,$; & 
$S'_1\,$&~$\mathbf{(1, 1)}_{Q35}\,$; &
$S\,$&~$\mathbf{(1, 1)}_{Q43}$ \\
\hline
$\Sigma_2\,$ & $L_2\,$ &~$\mathbf{(4, \bar 2)}_{Q12}\,$; &
$H_d\,$& ~$\mathbf{(1, 2)}_{Q24}\,$; &
$S_1\,$& ~$\mathbf{(1, 1)}_{Q35}\,$; &
$S'\,$ & ~$\mathbf{(1, 1)}_{Q43}$ \\
\hline
$\Sigma_3$ & $L_X\,$ & ~$\mathbf{(4, 1)}_{Q15}\,$; &
$H_u\,$ & ~$\mathbf{(1, \bar 2)}_{Q32}\,$; &
$R_3^d\,$ & ~$\mathbf{(\bar 4, 1)}_{Q41}\,$; &
$S_2\,$ & ~$\mathbf{(1, 1)}_{Q54}$ \\
\hline 
\end{tabular}
\end{center}
\end{table}

{}From the trilinear terms in the 6D bulk action,
we obtain the Yukawa terms
\begin{eqnarray}
{\cal S} &\!\!=&\!\! \int d^6 x \left[ \int d^2 \theta \ g_6 \left(
L_3 R_3^d H_d + L_2 R_3^d H'_d +
S H_d H_u + S' H'_d H_u  \right.\right. \nonumber \\
&& \qquad\qquad\qquad\qquad\qquad
 + S S_1 S_2 +
S' S_1' S_2 )
 + {\rm H. C.}\biggr]\, .
\end{eqnarray}

Similar to the above up-type partial unification model, 
$L_2$ can be considered as the left-handed fermions in
the second family in the SM, and the extra particles
$H_d'$, $L_X$, $S'$ and $S'_1$ can become very heavy after
the $U(1)'$ gauge symmetry is broken at the GUT scale.
Thus, after the $U(1)'$ gauge symmetry is broken,
we can have the relevant superpotential
\begin{eqnarray}
{\cal S} &\!\!=&\!\! \int d^6 x \left[ \int d^2 \theta \ g_6 \left(
L_3 R_3^d H_d  + S H_d H_u  + S S_1 S_2  \right)
 + {\rm H. C.}\right] .
\end{eqnarray}
giving rise to the unification of the gauge couplings with 
$\lambda$, $\kappa$ and $y_b$.

As a remark, to break the $SU(4) \times SU(2)_L \times U(1)_{I_{3R}}$
gauge symmetry down to the SM gauge symmetry by Higgs mechanism
in our models, on the 3-brane at $z=0$,
we emphasize that we introduce
the Higgs fields with the same quantum numbers as those of the right-handed
neutrino and its Hermitian conjugate.
Of course, if we did not specify the quantum numbers of the
Higgs fields, we can not distinguish the up-type partial unification and
down-type partial unification models.

\section{ 7D $SU(9)$ Models}
\setcounter{equation}{0}

In 6D $SU(9)$ models on $T^2/Z_6$ orbifold in previous Section,
unwanted fields remain in the zero mode spectrum after the orbifold projection.
When the models are constructed in 7D, the orbifold projection
is more powerful, and the unwanted fields can be projected out.

We consider the 7D space-time $M^4\times T^2/Z_6 \times S^1/Z_2$.
The coordinates are $x^{\mu}$ ($\mu = 0, 1, 2, 3$),
$x^5$, $x^6$ and $x^7$. Because $T^2$ is homeomorphic to $S^1\times S^1$,
we assume that the radii for the circles along the 
$x^5$, $x^6$ and $x^7$ directions
are $R_1$, $R_2$, and $R'$, respectively.
We define the complex
coordinate $z$ for $T^2$ and the real coordinate $y$ for $S^1$
\begin{eqnarray}
z \equiv{1\over 2} \left(x^5 + i x^6\right) , \quad y \equiv x^7 .
\end{eqnarray}

The $T^2/Z_6$ orbifold is defined in the Appendix A, and
the $S^1/Z_2$ orbifold is obtained from $S^1$ by moduloing the
equivalent class
\begin{equation}
\Gamma_S:~~~y\sim -y \,.
\end{equation}
There are two fixed points: $y=0$ and $y=\pi R'$.

The ${\cal N}=1$ supersymmetry in 7D has 16 supercharges and
 corresponds to the ${\cal N}=4$ supersymmetry in 4D,
thus, only the gauge multiplet can be introduced in the bulk.  This
multiplet can be decomposed under the 4D
 ${\cal N}=1$ supersymmetry into a vector
multiplet $V$ and three chiral multiplets $\Sigma_1$, $\Sigma_2$, and 
$\Sigma_3$ in the adjoint representation, where the fifth and sixth 
components of the gauge
field, $A_5$ and $A_6$ are contained in the lowest component of $\Sigma_1$,
and the seventh component of the gauge
field $A_7$ is contained in the lowest component of $\Sigma_2$.

For the bulk gauge group $G$, we write down the  bulk action 
in the Wess-Zumino gauge and 4D ${\cal N}=1$ supersymmetry
language~\cite{Marcus:1983wb}
\begin{eqnarray}
  {\cal S} &\!\!=&\!\! \int d^7 x \, \Biggl\{
  {\rm Tr} \Biggl[ \int d^2\theta \left( \frac{1}{4 k g^2} 
  {\cal W}^\alpha {\cal W}_\alpha + \frac{1}{k g^2} 
  \left( \Sigma_3 \partial_z \Sigma_2 + \Sigma_1 \partial_y \Sigma_3
   - {\sqrt{2}} \Sigma_1 
  [\Sigma_2, \Sigma_3] \right) \right) 
\nonumber\\
  &&
+ {\rm H.C.} \Biggr] 
 + \int d^4\theta \frac{1}{k g^2} {\rm Tr} \Biggl[ 
  (\frac1{\sqrt{2}} \partial_z^\dagger + \Sigma_1^\dagger) e^{-2V} 
  (-\frac1{\sqrt{2}} \partial_z + \Sigma_1) e^{2V}
 + \frac14 \partial_z^\dagger e^{-2V} \partial_z e^{2V}
\nonumber\\
  && + 
  (\frac1{\sqrt{2}} \partial_y + \Sigma_2^\dagger) e^{-2V} 
  (-\frac1{\sqrt{2}} \partial_y + \Sigma_2) e^{2V}
 + \frac14 \partial_y e^{-2V} \partial_y e^{2V}
+ {\Sigma_3}^\dagger e^{-2V} \Sigma_3 e^{2V} 
\Biggr] \Biggr\} \,.
\label{eq:t2z6action}
\end{eqnarray}
{}From above action, we obtain  
the transformations of vector multiplet 
\begin{eqnarray}
  V(x^{\mu}, ~\omega z, ~\omega^{-1} {\bar z},~y) &=& R \,
 V(x^{\mu}, ~z, ~{\bar z},~y) R^{-1} \,, \\
\label{TVtrans}
  \Sigma_1(x^{\mu}, ~\omega z, ~\omega^{-1} {\bar z},~y) &=& 
\omega^{-1} R \,
\Sigma_1(x^{\mu}, ~z, ~{\bar z},~y) R^{-1}\,, \\
\label{T1trans}
   \Sigma_2(x^{\mu}, ~\omega z, ~\omega^{-1} {\bar z},~y) &=& 
 R \, 
\Sigma_2(x^{\mu}, ~z, ~{\bar z},~y)  R^{-1}\,, \\
\label{T2trans}
 \Sigma_3(x^{\mu}, ~\omega z, ~\omega^{-1} {\bar z},~y)  &=& 
\omega R\,
\Sigma_3(x^{\mu}, ~z, ~{\bar z},~y) R^{-1}\,, \\
\label{T3trans}
  V(x^{\mu}, ~z, ~ {\bar z},\,-y) &=& P \,
 V(x^{\mu}, ~z, ~{\bar z},~y) P^{-1}\,, \\
\label{SVtrans}
  \Sigma_1(x^{\mu}, ~ z, ~ {\bar z},\,-y) &=& 
 P \,
\Sigma_1(x^{\mu}, ~z, ~{\bar z},~y) P^{-1}\,, \\
\label{S1trans-2}
   \Sigma_2(x^{\mu}, ~ z, ~ {\bar z},\,-y) &=& 
-  P \,
\Sigma_2(x^{\mu}, ~z, ~{\bar z},~y)  P^{-1}\,, \\
\label{S2trans-2}
 \Sigma_3(x^{\mu}, ~ z, ~ {\bar z},\,-y)  &=& 
- P \,
\Sigma_3(x^{\mu}, ~z, ~{\bar z},~y) P^{-1}\,, 
\label{S3trans-2}
\end{eqnarray}
where we introduce the non-trivial
$R$ and $P$ to break the bulk gauge group. 

To break the $SU(9)$ gauge symmetry, we choose
the following $9\times 9$ matrix representations for 
$R$ and $P$
\begin{eqnarray}
R &=& {\rm diag} \left(+1, +1, +1, +1,
 \omega^{n_1}, \omega^{n_1}, \omega^{n_1}, \omega^{n_2}, \omega^{n_2} \right), \\
P &=& {\rm diag} \left(+1, +1, +1, +1, +1, +1, -1, -1, +1\right),
\end{eqnarray}
where $n_1 \not= n_2 \not= 0$.
Then, we obtain
\begin{eqnarray}
 SU(9)/R &\!\!=&\!\! SU(4)\times SU(3)\times SU(2) \times U(1)^2 \,, \\
SU(9)/P &\!\!=&\!\! SU(7)\times SU(2) \times U(1) \,, \\
 SU(9)/\{R \cup P\}
&\!\!=&\!\! SU(4) \times SU(2)_L\times U(1)' \times U(1)_{\alpha}
\times U(1)_{\beta} \times U(1)_{\gamma}\,,
\end{eqnarray}
%
where the quotient $G/H$ denotes the commutant of $H$ in $G$.
The generators for the $U(1)' \times U(1)_{\alpha}
\times U(1)_{\beta} \times U(1)_{\gamma}$ gauge symmetry
are defined in Eqs.~(\ref{GU1P})-(\ref{GU1C2}).

The $Z_6\times Z_2$ transformation property for the decomposed components
of $V$ is 
\begin{equation}
V : \left(
\begin{array}{ccccc}
(1, +) & (\omega^{-n_1}, +) & (\omega^{-n_1}, -) & 
(\omega^{-n_2}, -) & (\omega^{-n_2}, +) \\
(\omega^{n_1}, +) & (1, +) & (1, -) & (\omega^{n_1-n_2}, -)
& (\omega^{n_1-n_2}, +) \\
(\omega^{n_1}, -) & (1, -) & (1, +) & (\omega^{n_1-n_2}, +)
& (\omega^{n_1-n_2}, -) \\
(\omega^{n_2}, -) & (\omega^{n_2-n_1}, -) & (\omega^{n_2-n_1}, +)
& (1, +) & (1, -) \\
(\omega^{n_2}, +) & (\omega^{n_2-n_1}, +) &
 (\omega^{n_2-n_1}, -) & (1, -) & (1, +) 
\end{array}
\right)  +  (1, +) \,,
\label{trans-1}
\end{equation}
and the transformation properties for the chiral multiplets, $\Sigma_i$, 
are also obtained from Eqs.~(\ref{TVtrans})-(\ref{S3trans-2}).
The zero modes transform as $(1,+)$.


Similar to the 6D up-type partial
unification model,  the  generator for the $U(1)_{I_{3R}}$
gauge symmetry is defined in Eq.~(\ref{U1B-L-U}).

We choose $n_1=5$ and $n_2=4$.
The zero modes from  the chiral multiplets 
$\Sigma_1$, $\Sigma_2$ and $\Sigma_3$ are given in
the Table \ref{Spectrum-UP-7D}.
It is easy to check that for  $L_3$, $S$, 
$S_1$, $S_2$ and $S_X$, the $U(1)_{I_{3R}}$ charges are
zero; for $H_u$ and $X_1$, the $U(1)_{I_{3R}}$ charges are
$1/2$; and for $H_d$, $R_3^u$ and $X_2$,
 the $U(1)_{I_{3R}}$ charges are $-1/2$.

\begin{table}[t]
\caption{The zero modes of the chiral multiplets 
$\Sigma_1$, $\Sigma_2$ and $\Sigma_3$
in the 7D orbifold $SU(9)$ model
with up-type partial unification.
\label{Spectrum-UP-7D}}
\begin{center}
\begin{tabular}{|c|c@{:}cc@{:}cc@{:}lc|}
\hline 
Chiral Fields & \multicolumn{7}{c|}{Zero Modes}  \\
\hline\hline
$\Sigma_1$ & $L_3\,$ &~$\mathbf{(4, \bar 2)}_{Q12}\,$; &
$X_1\,$&~$\mathbf{(1, 2)}_{Q25}\,$; &
$S\,$&~$\mathbf{(1, 1)}_{Q34}$ &
 \\
\hline
$\Sigma_2\,$ & $H_u\,$ &~$\mathbf{(1, 2)}_{Q23}\,$; &
$X_2\,$& ~$\mathbf{(1, \bar 2)}_{Q32}\,$; &
$S_1\,$& ~$\mathbf{(1, 1)}_{Q45}\,$; &
$S_X\,$:~$\mathbf{(1, 1)}_{Q54}$ \\
\hline
$\Sigma_3$ & $R_3^u\,$ & ~$\mathbf{(\bar 4, 1)}_{Q31}\,$; &
$H_d\,$ & ~$\mathbf{(1, \bar 2)}_{Q42}\,$; &
$S_2\,$ & ~$\mathbf{(1, 1)}_{Q53}\,$ &
 \\
\hline 
\end{tabular}
%
%
\end{center}
\end{table}

{}From the following trilinear term in the 7D bulk action
\begin{equation}
{\cal S} = \int d^7 x \left[\int d^2 \theta \frac1{k} {\rm Tr} 
\left(- {{\sqrt 2}} \,g\, \Sigma_1 [\Sigma_2,\Sigma_3]\right)
+ {\rm H.C.} \right],
\label{TriTerm-7D}
\end{equation}
we obtain the Yukawa terms
\begin{equation}
{\cal S} = \int d^7 x \left[ \int d^2 \theta \ g_7 \left(
L_3 R_3^u H_u + S H_d H_u  + S S_1 S_2 +
S_2 X_1 X_2 \right)
 + {\rm H. C.}\right].
\end{equation}

Because the $U(1)'$ gauge symmetry
can be broken at the GUT scale, the extra particles
$X_1$, $X_2$ and $S_X$ can become very heavy after 
it is broken. For simplicity, we do not consider
the $R$ symmetry here. 
To give the masses to $X_1$ and $X_2$, 
on the 3-brane at $z=0$ and $y=0$,
we introduce three 3-brane localized fields 
${\widetilde S}$, ${\widetilde X}_1$
and ${\widetilde X}_2$ with 
respectively quantum numbers 
$\mathbf{(1,  1)}_{{(-1, 0, 0, 0)}}$,
$\mathbf{(1,  2)}_{{(1, -1, 1, -9)}}$
and $\mathbf{(1,  2)}_{{(0, 2, 0, 0)}}$
under the $SU(4) \times SU(2)_L\times U(1)' \times U(1)_{\alpha}
\times U(1)_{\beta} \times U(1)_{\gamma}$ gauge symmetry.
After ${\widetilde S}$ gets VEV, 
the fields ($X_1$, ${\widetilde X}_1$) and
($X_2$, ${\widetilde X}_2$)  can obtain the 
vector-like masses through the following 3-brane localized
superpotential
\begin{equation}
W ~=~ {\widetilde S} X_1 {\widetilde X}_1
+ {\widetilde S} X_2 {\widetilde X}_2 \,.
\end{equation}
Similarly,  $S_X$ can also become very heavy.

In short, after the $U(1)'$ gauge symmetry is broken,
we can have the relevant superpotential
\begin{equation}
{\cal S} = \int d^7 x \left[ \int d^2 \theta \ g_7 \left(
L_3 R_3^u H_u  + S H_d H_u  + S S_1 S_2  \right)
 + {\rm H. C.}\right] .
\end{equation}

Similar to the previous Section,
the  down-type partial unification can be obtained just by flipping
the sign of $T_{I_{3R}}$.

\section{Numerical Studies for Coupling Unifications}
\setcounter{equation}{0}

\subsection{Predictions for the NMSSM Couplings}

As was pointed out, below the compactification scale we can have only the 
NMSSM-like particle
content in Higgs sector: $H_u$ and  $H_d$ the SM Higgs doublets and $S$ the SM singlet.  
The coupling $\lambda$ for $S H_u H_d$ is unified with the gauge couplings 
at the compactification scale.
When the bulk gauge symmetry is extended, extra SM singlets, $S_1$ and $S_2$,
can also be included in the massless modes below the compactification scale
and form a singlet trilinear coupling.
Therefore, when a bulk gauge symmetry with an enough large rank is considered, the 
NMSSM superpotential,
\begin{equation}
W_{\rm NMSSM}=\lambda H_uH_dS-\kappa SS_1S_2 \,,
\label{mm5}
\end{equation}
is naturally generated from the bulk gauge interaction.
%
%
Assuming that the compactification scale and gauge coupling unification scale ($M_{GUT}$)
are the same in the unified models, we  have the
 following condition,
\begin{equation}
g_1=g_2=g_3=\lambda=\kappa \,.
\label{mm7}
\end{equation}

\begin{figure}
\begin{center}
\includegraphics[viewport=20 15 278 220, clip, width=8cm]{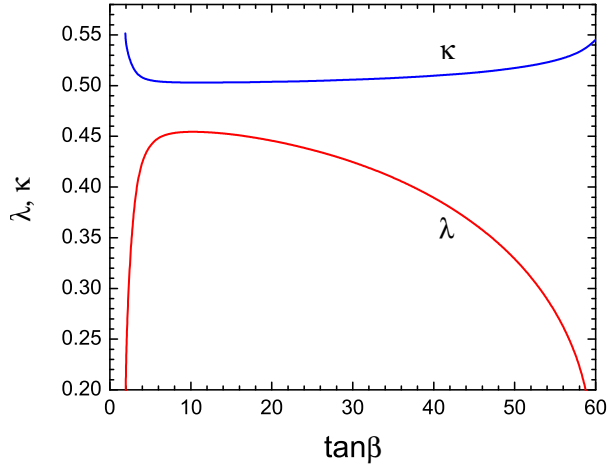}
\end{center}
\caption{ The Higgs trilinear couplings $\lambda$ and $\kappa$ 
(red and blue lines respectively) versus  the $\tan\beta$. 
We use top quark mass to be $178$ GeV.}
\label{fig1}
\end{figure}

Due to a crucial reduction of the number of the fundamental
parameters from the gauge-trilinear Higgs coupling unification, we are lead immediately
to a series of the very distinctive predictions 
(in absence of any large supersymmetric threshold corrections). 
Using the values of the electroweak parameters $\sin^2\theta_W=0.23120\pm0.00015$ and
$1/\alpha_{EM}=127.918\pm0.018$ in $\overline{MS}$ scheme 
at $M_Z$ scale~\cite{Eidelman:2004wy},
we can determine the unification scale and the unified coupling constant.
Then, evolving the remaining couplings from the unification scale
to the low energy, we predict $\kappa$ and $\lambda$ as functions of $\tan\beta$ 
(see Fig.~\ref{fig1}) where 
$\tan\beta \equiv \langle H_u^0 \rangle/\langle H_d^0 \rangle$.
In our numerical calculations, we have used the 
one-loop RGEs for  $\kappa$ and $\lambda$ and two-loop RGEs for Yukawa 
and gauge couplings~\cite{King:1995ys}
\begin{eqnarray}
\frac{d \alpha_{\lambda}}{dt} &\!\!=&\!\! 
\frac{\alpha_{\lambda}}{2\pi}(\alpha_{\kappa}+4\alpha_{\lambda}+3 \alpha_t + 3 \alpha_b + \alpha_\tau
- \frac{3}{5}\alpha_1 - 3\alpha_2)\,, \\
%
\frac{d \alpha_{\kappa}}{dt} &\!\!=&\!\!\frac{\alpha_{\kappa}}{2\pi}(3\alpha_{\kappa}+2\alpha_{\lambda})\,, \\
%
\frac{d \alpha_{t}}{dt} &\!\!=&\!\! \left[ \frac{d \alpha_{t}}{dt}\right]  _{\rm MSSM} 
+ \frac{\alpha_{t}}{2\pi}\left( \alpha_{\lambda} 
-\frac{1}{4\pi}\alpha_{\lambda}\left( 3\alpha_t+ 4\alpha_b + \alpha_\tau +
3\alpha_{\lambda}+\alpha_{\kappa}\right) \right) \,, \\
%
\frac{d \alpha_{b}}{dt} &\!\!=&\!\! \left[ \frac{d \alpha_{b}}{dt}\right]  _{\rm MSSM} 
+ \frac{\alpha_{b}}{2\pi}\left( \alpha_{\lambda} 
-\frac{1}{4\pi}\alpha_{\lambda}\left( 4\alpha_t+ 3\alpha_b +
3\alpha_{\lambda}+\alpha_{\kappa}\right) \right) \,, \\
%
\frac{d \alpha_{\tau}}{dt} &\!\!=&\!\! \left[ \frac{d \alpha_{\tau}}{dt}\right]  _{\rm MSSM} 
+ \frac{\alpha_{\tau}}{2\pi}\left( \alpha_{\lambda} 
-\frac{1}{4\pi}\alpha_{\lambda}\left( 3\alpha_t+ 3\alpha_\tau +
3\alpha_{\lambda}+\alpha_{\kappa}\right) \right) \,, \\
%
\frac{d \alpha_{2}}{dt} &\!\!=&\!\! \left[ \frac{d \alpha_{2}}{dt}\right] _{\rm MSSM}
+ \frac{\alpha_{2}^2}{8\pi^2}\left( -2\alpha_{\lambda}\right) \,, \\
%
\frac{d \alpha_{1}}{dt} &\!\!=&\!\!\left[ \frac{d \alpha_{1}}{dt}\right] _{\rm MSSM}
+ \frac{\alpha_{1}^2}{8\pi^2}\left( - \frac{6}{5}\alpha_{\lambda}\right) \,,
\end{eqnarray}
where $t$ is the log of renormalization scale,
$\alpha_i={g^2_i}/({4\pi})$, $\alpha_{t,b,\tau} = y_{t,b,\tau}^2/(4\pi)$,
$\alpha_\lambda = \lambda^2/(4\pi)$, $\alpha_\kappa = \kappa^2/(4\pi)$,
and  bracket $[\,]_{\rm MSSM}$ denotes 
the corresponding two-loop RGEs in the MSSM.
The $\lambda$ coupling depends on $\tan\beta$, while
the $\kappa$ coupling less depends on $\tan\beta$ and is predicted to be from
 0.5 to 0.55. For large $\tan\beta$, the $\lambda$ coupling at low energy
is sensitive to the running bottom-tau Yukawa couplings.
In Fig.~\ref{fig1}, we neglect the possible large supersymmetry
threshold corrections to bottom quark mass.

\begin{figure}
\begin{center}
\includegraphics[viewport=15 15 278 220, clip, width=8cm]{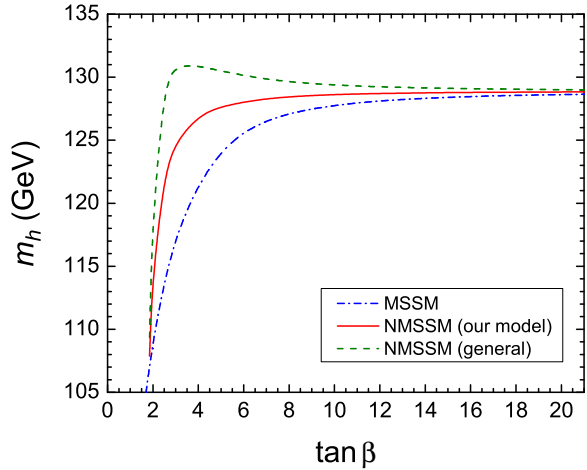}
\end{center}
\caption{ We present the upper bounds on the lightest CP-even Higgs 
mass in various models:
 green dashed line corresponds to the general NMSSM case, 
the blue dash-dotted line corresponds
to  the MSSM, and the red solid line corresponds to our model. }
\label{fig2}
\end{figure}

Since the value of $\lambda$ constant at low energy scale is predicted,
 we can calculate upper bound on the lightest CP-even neutral Higgs mass.
We use the following analytic formula
which includes full one-loop and 
the dominant two-loop top/stop corrections~\cite{Okada:1990vk}:
\begin{eqnarray}
m_h^2  & \!\!\simeq &\!\!   M_Z^2 \left ( \cos^2  2\beta +
\frac{2\lambda^2}{g^\prime{}^2+g_2^2} \, \sin^2 2\beta \right ) \left ( 1 -
\frac{3 \bar m_t^2}{8\pi^2 v^2} \, t \right ) \label{below1} \\
&+\!\!&\!\!   \frac{3 \bar m_t^4}{4\pi^2v^2} \,  \left (
\frac{1}{2} \, {X}_t + t + \frac{1}{16\pi^2} \left ( \frac{3}{2} \,
\frac{\bar m_t^2}{v^2} - 32\pi\alpha_3 \right ) ( {X}_t + t ) t \right ) , \nonumber
\label{mm11}
\end{eqnarray}
where $t=\log(M_{S}^2/m_t^2)$, $v=174$ GeV is the usual SM Higgs VEV, 
and 
\begin{eqnarray}
X_t =\frac{2 \tilde A_t^2}{M_{S}^2}\left (1-\frac{\tilde
A_t^2}{12 M_{S}^2}\right),
\label{mm12}
\end{eqnarray}
where $\tilde A_t=A_t-\mu\cot\beta$ is the top squark mixing parameter and $\mu$ is the
 supersymmetric Higgs mass parameter.
The supersymmetric scale, $M_{S}^2=(M^2_{\tilde t_1}+M^2_{\tilde t_2})/2$, 
 is the average of the two stop squared-masses. 
The $\bar m_t$ is the running
top quark mass at $\bar m_t$ in $\overline{MS}$ scheme.
In our calculations, we use the top quark pole mass 
$m_t=178$ GeV, and
 $M_{S}=1$ TeV. From the Eq.~(\ref{mm12}),
we can see that the maximal value for $X_t$ is $X_t=6$,
 and we use this value to calculate
 the lightest CP-even Higgs mass upper bounds
in various models presented in Fig.~\ref{fig2}.

\subsection{Predictions for Yukawa Coupling Constants}

Yukawa couplings for quarks and leptons in the third generation
can be unified with the gauge couplings at compactification scale
in the context of gauge-Higgs unification. 
The gauge-Yukawa (for top-bottom-tau) unification,
\begin{equation}
g_1 = g_2 = g_3 = y_t = y_b = y_\tau \,, \label{unif-1}
\end{equation}
is studied
in Ref.~\cite{Gogoladze:2003pp}.
The solid prediction of those coupling unification is $\tan\beta$.
As long as the unification condition, Eq.~(\ref{unif-1}), 
is satisfied within 5\% at unification scale,
the $\tan\beta$ is predicted as $\tan\beta = 52 \pm 1$
when the supersymmetry threshold corrections for tau mass is within a few percents.
The predictions of quark masses depend on the low energy 
supersymmetry threshold corrections.
Inputting the experimental data for the SM fermion masses,
$m_\tau = 1.777$ GeV and $\overline m_b (\overline m_b) = 4.26$ GeV (central values
from recent lattice calculation)
in $\overline{MS}$ scheme~\cite{Eidelman:2004wy},
the threshold corrections to the bottom quark mass should be less than several percents.
Since the finite corrections to the bottom quark mass is proportional to $\tan \beta$,
the threshold corrections,
\begin{equation}
\delta_b^{\rm finite} \simeq -\frac{\alpha_3}{3\pi} \frac{\mu M_{\tilde g} \tan \beta}{m_{\tilde b}^2}
+ \frac{\alpha_t}{8\pi} \frac{\mu A_t \tan\beta}{m_{\tilde t}^2}\,,
\end{equation}
could be 50\% when the superparticles' masses are about the same. However,
there exists the cancellation between gluino mass $M_{\tilde g}$ and 
trilinear scalar coupling for stop $A_t$.
Such cancellation is needed if we consider the 
top-bottom-tau Yukawa unification~\cite{Baer:2001yy}.
Although the detail prediction of top 
quark mass depends on the supersymmetric threshold corrections,
the top quark mass prediction is in good agreement with experiment.

\begin{figure}
\begin{center}
\includegraphics[viewport=10 15 278 220, clip, width=8cm]{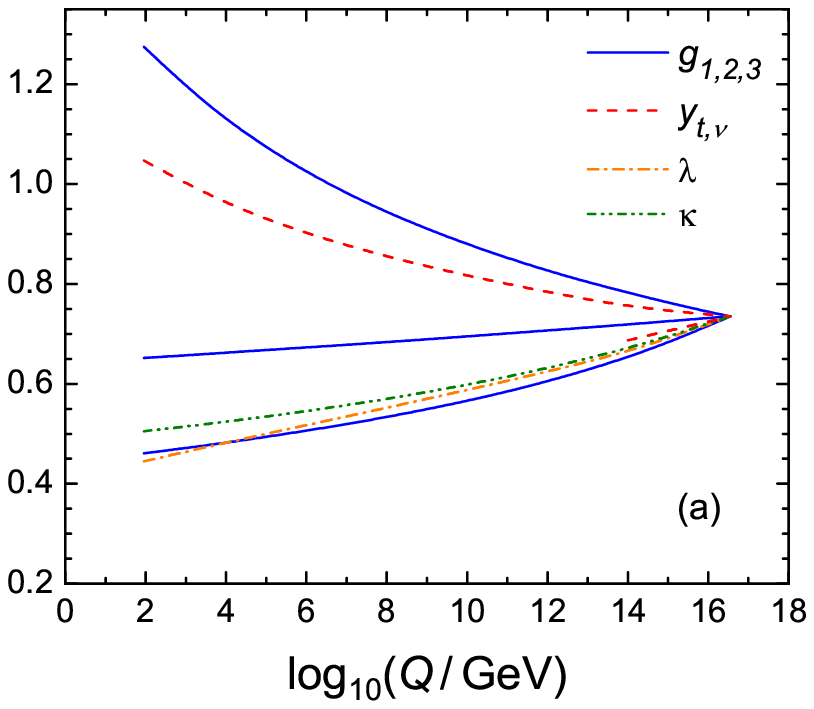}
\includegraphics[viewport=10 15 278 220, clip, width=8cm]{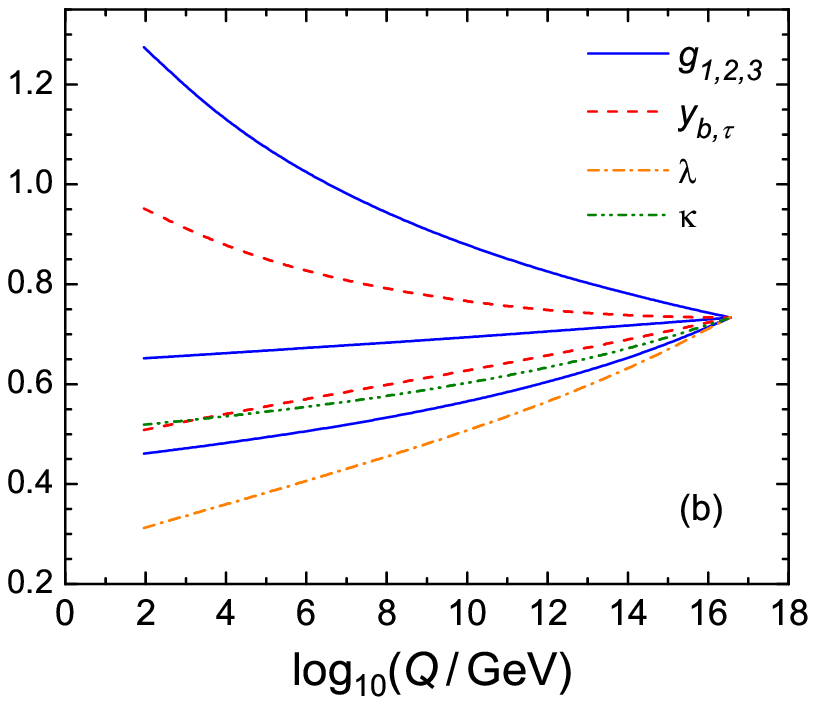}
\end{center}
\caption{
The unification of the gauge ($g_i$) and Yukawa ($y_f$) and Higgs 
($\lambda$ and $\kappa$) couplings.
Figure (a) shows up-type partial Yukawa unification,
 and figure (b) shows the down-type one.}
\label{fig3}
\end{figure}

We also present the evaluation of RGEs for the gauge couplings,
 Yukawa couplings for the third family, and Higgs trilinear couplings
$\kappa$ and $\lambda$ 
with unification condition at GUT scale in Fig.~\ref{fig3}. 
In those figures, we include the standard supersymmetric
threshold corrections at low energy by taking a single scale
$M_{SUSY}=M_Z$~\cite{Langacker:1992rq}.
Fig.~3(a) shows the coupling unification in the up-type partial unification.
In this case, the figure is shown for small $\tan\beta$ ($\tan\beta \simeq 2-15$)
and then the bottom and tau Yukawa couplings are small.
The neutrino Dirac Yukawa coupling evolution is also included above the right-handed 
neutrino Majorana mass scale, which is assumed to be $10^{14}$ GeV.
In Fig.~3(b), the down-type partial unification is shown.
In this case, $\tan\beta$ is large ($\tan\beta \simeq 51$), and  
the top and bottom Yukawa couplings are comparable.

\begin{figure}
\begin{center}
\includegraphics[viewport=15 15 278 220, clip, width=8cm]{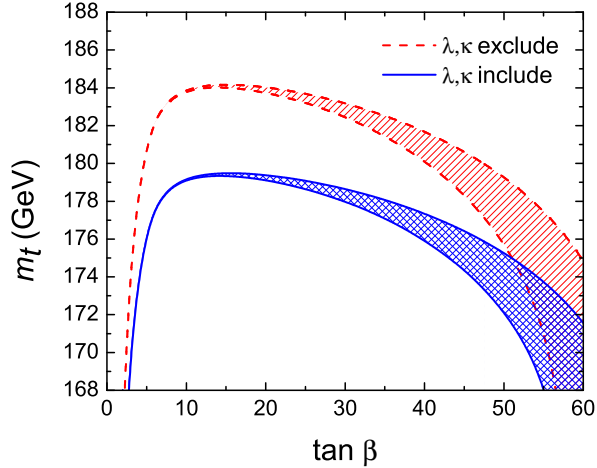}
\end{center}
\caption{Top quark mass prediction is given as the function of $\tan\beta$.
The red dashed lines are for the MSSM with unification of gauge and 
top quark Yukawa couplings.
The blue solid lines are given as the NMSSM with gauge, top quark Yukawa, 
and Higgs coupling unification.
For large $\tan\beta$, the top quark mass prediction is sensitive to
the input bottom quark running mass.
The shaded areas are shown for the bottom quark mass in the range,
$\overline m_b (\overline m_b) = 4.26 \pm 0.30$ GeV.
The bottom and tau Yukawa couplings can also be unified around $\tan\beta\simeq 50$.
}
\label{top-mass}
\end{figure}

Because the RGE for top quark Yukawa coupling, $\alpha_t$, is given in one-loop 
(including Dirac-neutrino Yukawa coupling),
\begin{equation}
\frac{d \alpha_t}{dt} = \frac{\alpha_t}{2\pi} \left(
6 \alpha_t + \alpha_b + \alpha_{\nu_\tau} + \alpha_\lambda -
\frac{13}{15} \alpha_1 - 3 \alpha_2 - \frac{16}{3} \alpha_3 \right),
\end{equation}
more coupling unification predicts the less top quark mass.
We show the prediction of top quark mass in Fig.~\ref{top-mass}.
In the figure, we assume that the couplings ($g_1$, $g_2$, and $y_t$ in the MSSM, and
$g_1$, $g_2$, $y_t$, $\lambda$, and $\kappa$ in the NMSSM) in 
$\overline{DR}$ scheme are unified at the GUT scale.
We use $\alpha_3 (M_Z) = 0.1187$ in $\overline{MS}$ scheme~\cite{Eidelman:2004wy}.
The top quark mass prediction in the NMSSM  (blue solid line)
is less than that in the MSSM (red dashed line).
The cut in the low $\tan\beta$ region in the figure is just coming from $\sin\beta$ factor 
in the top quark mass.
For large $\tan\beta$ ($\simeq 50)$, the 
bottom and tau Yukawa couplings can also be unified
(neglecting the possible large finite corrections to the bottom quark mass).
In this figure, we do not assume the bottom-tau unification at GUT scale, 
but we use the experimental data for the SM fermion masses,
$m_\tau = 1.777$ GeV and $\overline m_b (\overline m_b) = 4.26 \pm 0.30$ GeV 
in $\overline{MS}$ scheme.
In the large $\tan\beta$ region, the bottom quark threshold corrections are sensitive 
to the predicted top quark mass.
In the figure, the possible large supersymmetry threshold corrections for top quark mass 
are neglected.
When the supersymmetry mass parameters are set to be equal to a single scale $M_{SUSY}$,
the top quark mass prediction is insensitive to the $M_{SUSY}$ (when $M_{SUSY} \agt 200$ GeV). 
However, the decoupling type threshold corrections 
(squarks for only the first and second generations are heavy)
 may give 1 - 2\% level corrections 
to top quark mass.
Thus, the top quark mass prediction may have the uncertainties around $\pm 3$ GeV
due to the threshold corrections when the superpartners are heavy.

Without $\lambda$ coupling unification, the top quark mass prediction is 
larger than experimental value
for the middle range of $\tan\beta$.
In this case, the small $\tan\beta$ ($\alt 3$) is disfavored due to the
lightest Higgs mass upper bound, while the
large $\tan\beta$ ($\simeq 50$), {\it i.e.},
the  gauge and top-bottom-tau Yukawa unification,
is favored for the experimental value of the top quark pole mass.
On the other hand, in the case that $\lambda$ coupling is also unified to
the gauge and top quark Yukawa couplings
$\tan\beta \sim 5$
can be a solution (In this case, the lightest Higgs mass 
upper bound is around 128 GeV.).

Note that the $\kappa$ coupling unification is less important for the top quark
mass prediction
since the $\kappa$ gives only two-loop order corrections to the top quark Yukawa coupling.
The tau neutrino Dirac Yukawa coupling is also less important
since the top quark Yukawa prediction at low energy 
is insensitive to the GUT scale corrections, 
$\delta_t^{\rm GUT} \simeq y_{\nu_{\tau}}^2/(16\pi^2) \ln(M_{\rm GUT}/M_R)$,
where $M_R$ is the right-handed neutrino 
Majorana mass.

If we consider the gauge and only down-type Yukawa (bottom-tau) coupling unification,
the top quark mass is just an input parameter.
Since the top quark mass in the top-bottom-tau Yukawa unification
is in good agreement with the experimental data as an input parameter,
such an only down-type Yukawa and gauge coupling unification 
does not predict brand new physics.
In such a case, $\tan\beta$ is predicted around 50 in the same way as
the top-bottom-tau Yukawa and gauge coupling unification.
Conceptually, the up-type partial unification is 
very good since it explains why the  top quark mass is the heaviest among the
SM fermions naturally.

\section{Discussions and Conclusions}
\setcounter{equation}{0}


We have studied various possible coupling unifications in the gauge-Higgs unification
and gauge-Higgs-matter unification.
The gauge-Higgs unification in higher-dimensional supersymmetric models 
naturally leads to the unification of the gauge and Yukawa couplings.
In fact, the gauge and Yukawa coupling unification is naturally realized
when the grand unification is considered in higher-dimensional supersymmetric models.
One can also consider the gauge-Higgs unification in the context of the NMSSM.
The trilinear coupling $\lambda S H_u H_d$, which produces the $\mu$-term,
 can naturally be generated from the bulk interaction
in the simple choice of bulk gauge symmetries.
When large enough bulk gauge symmetries are considered,
the singlet trilinear coupling, $\kappa S S_1 S_2$, can also be included
in the bulk interaction.
The $\lambda$ and $\kappa$ couplings at low energy can be calculated 
by RGE evolution,
and thus we expect that the prediction will be tested in the future
collider experiments.

We constructed the $SU(8)$ and $SU(9)$ models,
which can realize the unification of 
the NMSSM Higgs trilinear couplings, Yukawa couplings, and gauge couplings.
We can select which Yukawa couplings (up-type
or down-type) are unified with the gauge and Higgs trilinear couplings
by the choice of bulk gauge symmetries and orbifold boundary conditions.
The up-type partial unification is very attractive
since it naturally explains why the top quark is the heaviest fermion.
The predictions can be in very good agreement with experimental observation
of the SM fermion masses.
Among them, there are two interesting regions allowed by the top quark mass.
One is the 
top-bottom-tau and gauge unification which predict $\tan\beta = 52 \pm 1$.
In this case, the bottom quark finite corrections should be 
at several percent level,
and thus the supersymmetry breaking parameters should be adjusted.
Another one is the $\lambda$ in the NMSSM, the top Yukawa and gauge coupling unification.
In the latter case, $\tan\beta$ can be around less than 5
in compatible with the current experimental lightest neutral Higgs mass upper bound.
The lightest Higgs mass upper bound
 is around 128 GeV when the supersymmetry breaking mass scale is 1 TeV,
and thus the models will be tested at the upcoming Large Hadron Collider.

\section*{Acknowledgments}

We would like to thank K. Tobe for useful discussions.
This research  was supported in part by the
National Science Foundation under Grants No.
PHY-0098791 (IG) and No.~PHY-0070928 (TL), by 
the Natural Sciences and Engineering Research Council of Canada (YM),
and by the Department of Energy 
Grants ~DE-FG02-04ER46140 and DE-FG02-04ER41306 (SN).

\appendix
\section{Gauge-Higgs Unification in 6D Supersymmetric Models on $T^2/Z_6$ Orbifold}
\setcounter{equation}{0}

We introduce the gauge-Higgs unification in 6D supersymmetric models.
%
We consider the 6D space-time which can be factorized into a product of the 
ordinary 4D Minkowski space-time $M^4$, and the torus $T^2$.
The corresponding
coordinates for the space-time are $x^{\mu}$ ($\mu = 0, 1, 2, 3$),
$ x^5$ and $ x^6$. 
The radii for the circles along the $x^5 $ and $x^6 $ directions are
$R_1$ and $R_2$, respectively. We define the
complex coordinate
\begin{eqnarray}
z \equiv{1\over 2} \left(x^5 + i x^6\right) \,.
\end{eqnarray}
In the complex coordinate,
the torus $T^2$ can be defined by $C^1$ moduloing the
equivalent classes: 
\begin{eqnarray}
z \sim z+ \pi R_1 \,, \quad z \sim z +  \pi R_2 e^{{\rm i}\theta} \,.
\end{eqnarray}
To define $T^2/Z_6$ orbifold, we require that $R_1=R_2\equiv R$
and $\theta = \pi/3$.
The $T^2/Z_6$ orbifold is obtained from $T^2$ by moduloing the
equivalent class
\begin{eqnarray}
\Gamma_T: \ \ z \sim \omega  z \,,
\end{eqnarray}
where $\omega =e^{{\rm i}\pi/3} $. There is one $Z_6$ fixed point
$z=0$, two $Z_3$ fixed points: 
 $z=\pi R e^{{\rm i}\pi/6}/{\sqrt 3}$ and
$z=2 \pi R e^{{\rm i}\pi/6}/{\sqrt 3}$, and three $Z_2$ fixed points:
$z=\sqrt 3 \pi R e^{{\rm i}\pi/6}/2$, $z=\pi R/2$ and $z= \pi R e^{{\rm i}\pi/3}/2$.


The ${\cal N} = (1, 1)$ supersymmetry in 6D
has 16 supercharges and
 corresponds to the ${\cal N}=4$ supersymmetry in 4D,
thus, only the gauge multiplet can be introduced in the bulk.  This
multiplet can be decomposed under the 4D
 ${\cal N}=1$ supersymmetry into a vector
multiplet $V$ and three chiral multiplets $\Sigma_1$, $\Sigma_2$, and 
$\Sigma_3$ in the adjoint representation, where the fifth and sixth 
components of the gauge
field, $A_5$ and $A_6$, are contained in the lowest component of $\Sigma_1$.

We write down the  bulk action 
in the Wess-Zumino gauge and 4D ${\cal N}=1$ supersymmetry
language~\cite{Marcus:1983wb},
\begin{eqnarray}
  {\cal S} &=& \int d^6 x \Biggl\{
  {\rm Tr} \Biggl[ \int d^2\theta \left( \frac{1}{4 k g^2} 
  {\cal W}^\alpha {\cal W}_\alpha + \frac{1}{k g^2} 
  \left( \Sigma_3 \partial \Sigma_2   - {\sqrt{2}} \Sigma_1 
  [\Sigma_2, \Sigma_3] \right) \right) + {\rm H.C.} \Biggr] 
\nonumber\\
  && + \int d^4\theta \frac{1}{k g^2} {\rm Tr} \Biggl[ 
  (\frac1{\sqrt{2}} \partial_z^\dagger + \Sigma_1^\dagger) e^{-2V} 
  (-\frac1{\sqrt{2}} \partial_z + \Sigma_1) e^{2V}
 + \frac14 \partial_z^\dagger e^{-2V} \partial_z e^{2V} \Biggr]
\nonumber\\
&&+ \int d^4\theta \frac{1}{k g^2} {\rm Tr} \Biggl[
   \Sigma_2^\dagger e^{-2V} \Sigma_2  e^{2V}
  + {\Sigma_3}^\dagger e^{-2V} \Sigma_3 e^{2V} 
\Biggr] \Biggr\}\,.
\label{eq:t2z6action-6}
\end{eqnarray}
The above action is invariant under the following orbifold transformation conditions:
\begin{eqnarray}
  V(x^{\mu}, ~\omega z, ~\omega^{-1} {\bar z}) &=& R\ 
 V(x^{\mu}, ~z, ~{\bar z}) R^{-1} \,, 
\label{Vtrans} \\
  \Sigma_1(x^{\mu}, ~\omega z, ~\omega^{-1} {\bar z}) &=& 
\omega^{-1}\, R\ 
\Sigma_1(x^{\mu}, ~z, ~{\bar z}) R^{-1} \,,
\label{S1trans} \\
   \Sigma_2(x^{\mu}, ~\omega z, ~\omega^{-1} {\bar z}) &=& 
\omega^{-1-m}\, R\  
\Sigma_2(x^{\mu}, ~z, ~{\bar z})  R^{-1} \,,
\label{S2trans} \\
 \Sigma_3(x^{\mu}, ~\omega z, ~\omega^{-1} {\bar z})  &=& 
\omega^{2+m}\, R \ 
\Sigma_3(x^{\mu}, ~z, ~{\bar z}) R^{-1} \,,
\label{S3trans}
\end{eqnarray}
where we introduce the non-trivial action on the gauge space, $R$,
which can break the bulk gauge group $G$ down to $H$ on the $Z_6$ fixed points.
%
To keep the 4D ${\cal N}=1$ supersymmetry, we 
need $m=0$ or $m=1$~\cite{Li:2003ee}.

Under the orbifold conditions, any bulk fields have $Z_6$ charges,
such as
\begin{equation}
\phi(x^\mu, \omega z, \omega^{-1}\bar z) = \omega^a \, \phi(x^\mu, z, \bar z)\,.
\end{equation}
 The bulk field with $a\neq0$
vanishes at the fixed point, and then does not have massless mode.
On the other hand, the bulk field with $a=0$ contains zero-mode in 4D.
By introducing non-trivial $R$, the
bulk vector multiplet is decomposed as the following in the matrix presentation:
\begin{equation}
V = \left( 
\begin{array}{cccc}
V^{(1)} & V^{(12)} & \cdots & V^{(1n)} \\ 
V^{(21)} & V^{(2)} & \cdots & V^{(2n)} \\ 
\vdots & \vdots & \ddots & \vdots \\ 
V^{(n1)} & V^{(n2)} & \cdots & V^{(n)} 
\end{array}
\right), \qquad
\Sigma_i = \left( 
\begin{array}{cccc}
\Sigma_i^{(1)} & \Sigma_i^{(12)} & \cdots & \Sigma_i^{(1n)} \\ 
\Sigma_i^{(21)} & \Sigma_i^{(2)} & \cdots & \Sigma_i^{(2n)} \\ 
\vdots & \vdots & \ddots & \vdots \\ 
\Sigma_i^{(n1)} & \Sigma_i^{(n2)} & \cdots & \Sigma_i^{(n)} 
\end{array}
\right),
\end{equation}
where each decomposed component can have different $Z_6$ charges.
By definition, $V^{(i)}$'s have charge 0, and
correspond to vector multiplet for gauge group $H$, while
$V^{(ij)}$'s, which correspond to $G/H$, do not have zero-modes.
On the other hand, due to the conditions in Eqs.~(\ref{S1trans}-\ref{S3trans}),
$\Sigma_i^{(i)}$'s do not contain massless modes unless $m = +1$, $-2$,
and some of the $\Sigma_i^{(ij)}$ can have massless modes.
The off-diagonal components in the decomposition correspond to
the  bifundamental representations under $H$ when $G=SU(N)$,
and then the zero-modes in $\Sigma_i^{(ij)}$ can be identified as the Higgs fields.
Thus, the gauge and Higgs unification is realized naturally.
Furthermore, the bulk action, Eq.~(\ref{eq:t2z6action-6}),
includes the trilinear couplings for the massless modes of $\Sigma_i^{(ij)}$,
\begin{equation}
{\rm Tr}\,\Sigma_1[\Sigma_2,\Sigma_3]= 
\Sigma_1^{(ij)} \left( \Sigma_2^{(jk)}\Sigma_3^{(ki)} - \Sigma_3^{(jk)}\Sigma_2^{(ki)} \right).
\end{equation}
Therefore, the trilinear coupling constants 
for the normalized scalar fields, $\sqrt2 \Sigma_i^{(ij)}$, 
are the same as the gauge coupling constants
in the conventional definition ($k=1/2$).

\end{document}